\documentclass[12pt]{article}
\usepackage{authblk}
\usepackage[T1]{fontenc}
\usepackage[utf8]{inputenc}
\usepackage{lmodern}
\usepackage[numbers]{natbib}
\usepackage{overpic}
\usepackage{amsmath}
\usepackage{amssymb}
\usepackage[english]{babel}
\usepackage[autostyle]{csquotes}
\usepackage{authblk}
\usepackage[usenames,dvipsnames]{color} 
\usepackage{graphicx}
\usepackage{geometry}
\usepackage[FIGTOPCAP]{subfigure}
\usepackage{units}
\usepackage{tikz}
\usetikzlibrary{calc} 

\title{A growth model driven by curvature reproduces geometric features of arboreal termite nests}
\author[1]{G. Facchini}
\author[2]{A. Lazarescu}
\author[1]{A. Perna}
\author[3]{S. Douady}

\affil[1]{Life Sciences Department, University of Roehampton, London, UK}
\affil[2]{Institut de Recherche en Mathématique et Physique, UC Louvain, Louvain-la-Neuve, Belgium}
\affil[3]{Laboratoire Matière et Systèmes Complexe, Université Paris Diderot, Paris, France}
\date{}

\begin{document}
\maketitle
\abstract
We present a simple three-dimensional model to describe the autonomous expansion of a substrate which grows driven by the local mean curvature of its surface. 
The model aims to reproduce the nest construction process in arboreal \textit{Nasutitermes} termites, whose cooperation may similarly be mediated by the shape of the structure they are walking on, for example focusing the building activity of termites where local mean curvature is high. 
We adopt a phase-field model where the nest is described by one continuous scalar field and its growth is governed by a single nonlinear equation with one adjustable parameter $d$. When $d$ is large enough the equation is linearly unstable and fairly reproduces a growth process where the initial walls expand, branch and merge, while progressively invading all the available space, which is consistent with the intricate structures of real nests. 
Interestingly, the linear problem associated to our growth equation is analogous to the buckling of a thin elastic plate under symmetric in-plane compression which is also known to produce rich pattern through non linear and secondary instabilities.
We validated our model by collecting nests of two species of arboreal \textit{Nasutitermes} from the field and imaging their structure with a micro-CT scanner. We found a strong resemblance between real and simulated nests, characterised by the emergence of a characteristic length-scale and by the abundance of saddle-shaped surfaces with zero-mean curvature which validates the choice of the driving mechanism of our growth model.

\section{Introduction}
Self-organised growth phenomena are ubiquitous in physics, societies and biology, with examples ranging from dunes and ripples in the sand, to crystals and cities, to developing plants and embryos (see \cite{bourgine2010morphogenesis} for a review).
All such phenomena are associated with autonomous processes in which a substrate increases its size while simultaneously developing a characteristic shape. 

In some systems growth is driven directly by volumetric expansion. This process is common for instance in living tissues, where cell proliferation can happen both at the surface and inside the volume. 
In other systems, growth is driven mainly by accretion processes, whereby new material is added or removed only at the surface of the growing object (crystals and concretions, but also corals and shells are example of this latter type of growth)\cite{thompson1942growth}. 
As a consequence, accretive growth is inevitably related to the dynamics of the interfaces between the growing substrate and the external environment.
Importantly the shape of the interface can contribute to focus the growth-driving quantity at specific positions thus causing the onset of positive feedback loops that produce fingering and tip-splitting. The growth-driving quantity itself can be different in different systems, for example it could be pressure, as in the seminal work of \cite{Saffman1958} - where a first fluid invades a more viscous one -, or it could be the concentration of a protein as in the more recent work of \cite{Clement2012}, aimed at modelling the branching morphogenesis of lungs.
At small length scales negative feedback (viscous dissipation, capillarity) stabilizes the system and fixes the typical scale of the final pattern \citep{Pelce2000}. Growth is then often a self-sustained process where the form of the substrate is the main ingredient as well as the main outcome of the growth process. 

A particular example of self-organised accretive growth is nest building by social insects. At first sight, collective nest building by social insects may look different from the examples above because the transport and deposition of building material is mediated by active agents - the insects - that are capable of multiple regulations of activity, and of complex movement patterns. 
In fact, there is extensive evidence that nest building can be influenced by a number of factors, including insects responding to environmental gradients \cite{Hangartner1969}, crowding \cite{Toffin2009}, the insects using their own body as a template \citep{Khuong2016}, and individual abilities such as proprioception \cite{bardunias2009dead} (see \cite{perna2017social} for a review).
Yet, while insects movement itself can follow complex rules, the decisions that insects take of adding or removing pellets at a particular place obey simple ``stigmergic'' principles whereby the local configuration of the environment is the main drive that determines the probability of deposition and collection of material \citep{Grasse1959,Theraulaz1999}. 
When insect density is high the effect of such stigmergic regulations is likely to dominate over the other factors and the nest growth resembles an accretive growth.
In this continuum approach the construction process should be described by a Turing-like system of coupled equations, one for each of the interacting fields - for example density of agents, pheromones, $CO_2$ concentration etc. - which are constrained by the shape of the structure under construction. 
In particular we expect the spatial boundaries to affect not only the spatial distribution but also the interactions among the different fields at play.
This suggests to consider an alternative and simplified approach where only such a spatial constraints are retained and the nest grows by itself as a function of its own shape, similarly to a crystal or a cell's tissue.

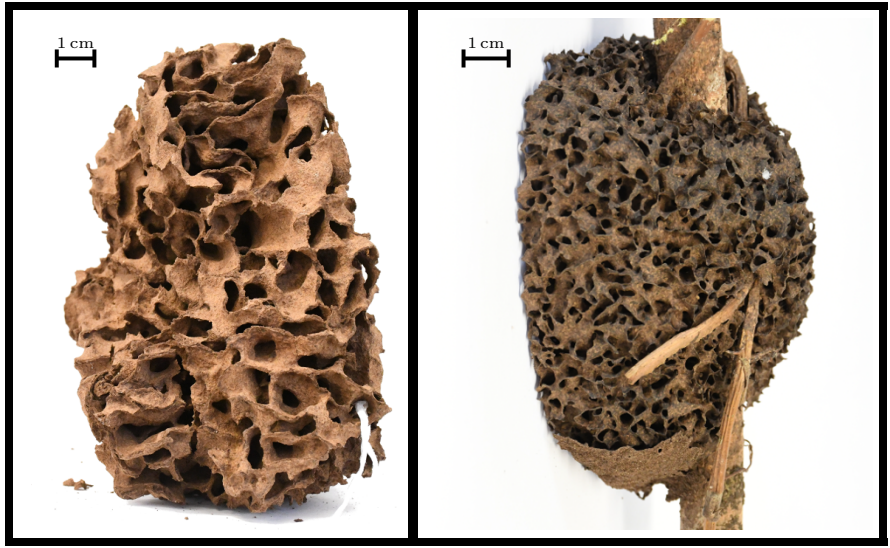
\begin{figure}
\centering
\noindent
\def\sH{.45}			
\subfigure
{
\textcolor{black}{\fboxrule=3pt\fbox{
\begin{overpic}[height=\sH\linewidth]{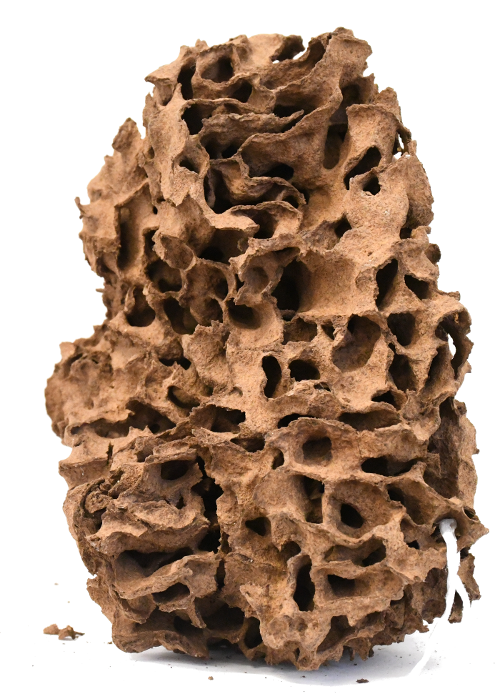}
\put (3,90) {\begin{tikzpicture}
\draw[line width =1.5,-,black] (0,0) -- (.5,0);
\draw[line width =1.5,-,black] (0,.12) -- (0,-.12);
\draw[line width =1.5,-,black] (.5,.12) -- (.5,-.12);
\node[] at (.25,0.2) {\tiny \unit[1]{cm}};
\end{tikzpicture}}
\end{overpic}}}
}
\hspace{-1.5em}
\subfigure
{
\textcolor{black}{\fboxrule=3pt\fbox{
\begin{overpic}[height=\sH\linewidth]{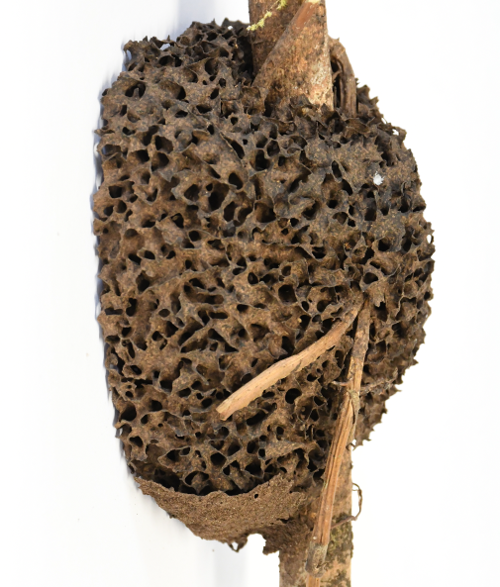}
\put (4,90) {\begin{tikzpicture}
\draw[line width =1.5,-,black] (0,0) -- (.6,0);
\draw[line width =1.5,-,black] (0,.12) -- (0,-.12);
\draw[line width =1.5,-,black] (.6,.12) -- (.6,-.12);
\node[] at (.3,0.2) {\tiny \unit[1]{cm}};
\end{tikzpicture}}
\end{overpic}}}
}
\caption{\label{fig:nasuti_nests} Fragments of nests built by \textit{Nasutitermes walkeri} (left) and \textit{Nasutitermes ephratae} (right).}
\end{figure}

Here, we focus in particular on modelling the nests built by arboreal termites of the genus \textit{Nasutitermes}.
These are lightweight - but resistant - structures that are usually built up on trees, typically around a branch or a tree fork.
Their structure is often isotropic and very homogeneous, to the point that different part of the nest can hardly been distinguished;
galleries show a typical length-scale and there are no chambers or other specialised structures (see Fig. \ref{fig:nasuti_nests}). 
As a consequence these nests resemble a continuum substrate with a coherent morphology, which makes them particularly suitable to be treated as the result of an accretive growth.
Our task is to identify a relevant and minimal ``stigmergic function'' that summarizes the results of the interactions of termites with the building material (i.e. the local rules of growth and remodelling of the built structure) and which is sufficient to produce structures comparable to the observed ones. 

The actual construction process is difficult to observe, as termites concentrate their building activity in short and unpredictable bursts.
The only available descriptions of termite behaviour during nest expansions come from the laboratory experiments of \cite{Jones1979} (from the same author see also \cite{Jones1977}).
Jones observed that construction always happens by deposition of material on the edge of existing walls, and that these edges progressively bend, split, and merge in order to form an intricate matrix, where the local surface of the walls often resembles to a saddle.  
These observations suggest that the local curvature of the wall surface focuses the growth activity. 
The idea that curvature might guide termite aggregation and building behaviour was investigated much more recently in experiments by \cite{Calovi2019} who showed the existence of a positive correlation between termites activity and surface curvature in \textit{Macrotermes michaelseni} termites (an African termite species not too closely related to \textit{Nasutitermes}). 
The observation that termite activity is driven by curvature does not exclude that cement pheromones or inter-individual interactions \citep{Fouquet2014,Green2017} are involved in the nest construction. 
Simply we suggest that either these quantities are directly affected by nest curvature, or they co-vary with it.

Consistently with these observations, we develop a model of surface evolution that depends directly on surface curvature. The model is reminiscent of gradient growth processes, where the local curvature of the surface concentrates gradients and fuels instabilities. 
In these processes, the instability is strong and results in separate branches that compete together \citep{Lajeunesse2000}, and cannot fuse \citep{Couder2005}. Here the nests surfaces are tightly interconnected so we have to find a new way of writing this growth, allowing both separation and fusion.

In section \ref{sec:model} we derive our minimal model and perform the linear analysis of our growth equation.
In section  \ref{sec:numerical} we briefly describe the numerical tools that allow to solve the full nonlinear equation and in section \ref{sec:results} we present the results of numerical simulations.
In section \ref{sec:comparison} we compare our simulated nest to the tomographies of real nest fragments collected in the field.
Finally in section \ref{sec:discussion} we discuss the results of the present study and illustrate the future applications of our model.  

\section{Growth model}\label{sec:model}
A large set of growth phenomena can be identified with the dynamics at the interface that divides the growing substrate from the medium that supports the growth.
One is then confronted with a two-phases problem where boundary conditions change in time and take the form of a curve or a surface, depending on whether the geometry of the problem is highly confined in one direction or fully three dimensional like our nests.
Whenever this curve or surface largely deviates from a planar front, it is quite challenging to treat the problem analytically, and predictions can be done only locally. Implementing a time evolving boundary condition in three dimensions is also difficult and would certainly require additional computing resources compared to a regular domain.
As an alternative approach, here we adopt a phase-field model \citep{Fix1983}, or diffuse interface method, which consists in replacing the substrate and the medium with a continuous phase endowed with a scalar field $f$ which is defined everywhere in the considered volume and determines the presence of the first or the second entity according to a threshold value. 
The frontier between the two phases can then be easily retrieved as an iso-contour of $f$ in the considered volume.
As long as $f$ is continuous, the geometry of the frontier can be also obtained from $f$ via differentiation \citep{Goldman2005}, the growth model is then extremely simple and consists of only one differential equation for the field $f$.
The growth equation should be nonlinear because we want the growth process to saturate at some point, and we also want to impose that new walls are only built adjacent to pre-existing ones, i.e. we want to describe an accretion process.
We define our scalar field $f$ in the interval $[0,1]$, and assume that the actual surface of the walls corresponds to the iso-surface $f=f_0$, for example $f_0=0.5$. 
We will then say that a point $\boldsymbol{x}$ is inside the wall whenever $f({\boldsymbol{x}})>f_0$ and outside otherwise.
If the nest growth is a function of its own shape, the growth equation must take the form:
\begin{equation}\label{eq:f_A}
\frac{\partial f}{\partial t} = \mathcal{A}(f)
\end{equation}
where the operator $\mathcal{A}$ is meant to mimic the main features of the growth process as they are inferred from nest observations, and we construct it as follows:
\begin{equation}\label{eq:A_components}
\mathcal{A}(f)=-f^{\alpha}(1-f)^{\beta}\cdot d(\nabla\cdot\boldsymbol{\hat{n}})-\Delta(\nabla\cdot\boldsymbol{\hat{n}})
\end{equation}
where the constant $d$ is strictly positive and the equation is presented in a non dimensional form. 
The unit vector $\boldsymbol{\hat{n}}$ is the normal to the local iso-surface and is given by 
$\boldsymbol{\hat{n}}=\nabla f/|\nabla f|$, for example ${\hat{n}}(\boldsymbol{x}_0)=\left(\nabla f/|\nabla f|\right)|_{\boldsymbol{x}_0}$ will be the normal to the iso-surface $f(\boldsymbol{x})=f(\boldsymbol{x}_0)$ 
in the point $\boldsymbol{x}_0$.

The first term of the equation represents the actual growth and can be decomposed in two factors: 
a differential operator $-d(\nabla\cdot\boldsymbol{\hat{n}})$ and a non linear kernel $f^{\alpha}(1-f)^{\beta}$. 
The first factor is precisely the mean curvature of the local iso-surface times a constant, and is intended to mimic how local curvature enhances the building or digging activity: the field $f$ increaseas (decreases) whenever the local curvature is positive (negative). 
The second factor vanishes at $f=0,1$ and is maximum at $f=f_0=\alpha/(\alpha+\beta)$ and is meant to damp any variation of $f$ far from the isosurface $f=f_0$ which is the only accessible space for our walking agents: there can't be any growth neither in the empty space far from the pre-existent nest, nor deep inside a pre-existent wall.

The second term in equation (\ref{eq:A_components}) is proportional to the diffusion of the mean curvature, which consequently compensates the growth term whenever the local curvature is too high.
Technically this is a dissipative term which provides a cutoff at the highest wavelength and is necessary to produce a meaningful pattern as is highlighted below in the linear analysis. Phenomenologically there are several reasons to introduce this term.
On one side, the growth of a wall happens (field and experimental observations) by the sequential deposition of faecal pellets and soil/sands particles of finite size, thus we can expect the existence of a minimum scale in the spatial structure. 
On the other side one can expect that, while ruled, the contributions of thousands of agents must sum up with some
noise whose average effect likely produces some diffusion. 
Finally an explicit justification can be found in the experimental observations of \cite{Jones1980} which include smoothing process among the tasks performed by individual workers.

Now that a growth equation is defined we want to make a further approximation.
In fact the curvature operator $\nabla\cdot\boldsymbol{\hat{n}}$  is highly non-analytic (and nonlinear) while we would like our equation to be as simple as possible for both analytical and numerical purposes. 
In the hypothesis that $|\nabla f|$ is almost constant in the nearby of $f=0.5$, one can approximate $\nabla\cdot\boldsymbol{\hat{n}}\propto \Delta f$.
In addition we take $\alpha=\beta=1$, which fixes the maximum of the nonlinear kernel and the contour of the actual nest at $f_0=0.5$. 
Thus, our simplified growth equation reads: 
\begin{equation}\label{eq:growth_0}
\frac{\partial f}{\partial t}=-f(1-f)d\Delta f-\Delta^2f,
\end{equation}
where $\Delta$ is the standard laplacian or diffusion operator and $\Delta^2$ is the bilaplacian operator  which commonly enters in elasticity problems \citep{Timoshenko1970} and was recently included in a phenomenological model for vegetation patterns \citep{Reynes2019}.
One observes that $\Delta f$ appears here with the opposite sign of a common diffusion equation, that is the first term in equation (\ref{eq:growth_0}) is anti-diffusive, while the second term is a diffusive term of higher order or a hyper-diffusion. 
Note that the difference in the derivative order of the two operators is the key ingredient for pattern formation as it will be shown below in the linear analysis.
Finally one may remark that equation (\ref{eq:growth_0}) does not conserve the total mass (or volume) because of the non linear pre-factor $f(1-f)$. 
We stress that these two elements are necessary to produce an organized structure like our termite nests as already observed by \cite{Deneubourg1977}. 
\subsection{Linear analysis}\label{subsec:linear}
In its simplified form (\ref{eq:growth_0}) the growth equation can be easily linearised around $f_0=0.5$ and solved in Fourier space with solutions of the form $f=f_0+\hat{f}\exp{[i\boldsymbol{k}\cdot\boldsymbol{x} + \sigma t]}$
which give the following dispersion relation:
\begin{equation}\label{eq:disp_relation}
\sigma(k)=\tilde{f}dk^2-k^4
\end{equation}
where for simplicity we wrote $|\boldsymbol{k}|$ and $\tilde{f}=f_0(1-f_0)=0.25$. 
The dispersion relation is shown in Fig. S1.
The parameter $d$ is strictly positive which means that $\sigma(k)$ changes sign at $k_0=(\tilde{f}d)^{1/2}$ and has a local maximum at $k_{max}=(\tilde{f}d/2)^{1/2}$: equation (\ref{eq:growth_0}) is linearly unstable with the most unstable length at $\lambda_{max}=2\pi/k_{max}$ while all the small scale disturbances beyond the cutoff length $\lambda_0=2\pi/k_{0}$ are damped.
Note that $k_0>0$ independently of $d$, thus the system is formally always unstable. 
Nonetheless a threshold is fixed by the smallest $k$ which can appear in a given domain of size $L$ which is $k=2\pi/L$. The instability threshold $d_c$ is then fixed by $L$ the size of the domain, with $d_c=(2\pi/L)^2/\tilde{f}$ being the marginal condition. 
Interestingly in two dimensions the same eigenvalues problem arises when studying the buckling modes of a thin plate under symmetric in-plane compression \citep{Timoshenko1962}.
The stationary two-dimensional version of equation (\ref{eq:growth_0}) coincides with a particular case of F\"oppl–von K\'arm\'an equations and were discovered already more than one century ago \citep{Foppl1907}. 
Nonetheless a very similar problem has gathered new attention in the last decades to explain delamination patterns occurring in multi-layered material where the coating material does not have the same elastic properties of the underlying substrate \citep{Ortiz1994,Bowden1998,Audoly1999}. 
Notably a similar mechanism was also invoked to explain the formation of Miura-ori folding patterns that approximately appears in nature, for example in insects wings and blooming leaves \citep{Mahadevan2005}.
In most cases emerging patterns significantly differ from the planar solution (i.e. the solution we chose at the beginning of this section) and likely arise from secondary instabilities \citep{Audoly2008} that rather select a superposition of planar waves which reflects the invariance of their orientation at the instability onset.

\section{Numerical simulations}\label{sec:numerical}
We simulate equation (\ref{eq:growth_0}) on a cubic grid with a finite difference code written in \textit{Python} and automatically parallelised with the \textit{Numba} compiler. 
The largest domain we considered is 256 x 256 x 256 where a growth simulation can be accomplished within about 5000 cpu hours. 
Both laplacian and bilaplacian operators are implemented with a first order explicit Euler scheme.   
Two different boundary conditions were considered: in the first configuration boundary conditions are triply periodic while in the second one lateral faces of the simulation box are still periodic but Dirichlet boundary conditions (i.e. $f=const$) are imposed on bottom and top faces. 
Initial conditions are also varied. 
In the simplest case the whole simulation domain is initiated with white noise, mainly to characterize the long term (or quasi-stationary) solutions of the equation within the bulk. 
Alternatively we initiate simulations with part of the volume copied from a previous simulation where a distinct pattern has already appeared, while the remaining part of the domain is set to zero. In this case we also forbid any evolution for all the points that are initially above a certain threshold which maintains the initial seed unchanged. 
This configuration is intended to mimic the expansion process of a pre-existing nest, or a new one which starts from an arbitrary shaped support (in the following we will refer to that as a nest-like seed).

\section{Results}\label{sec:results}
\begin{figure}
\centering
\subfigure{\includegraphics[height=0.16\linewidth]{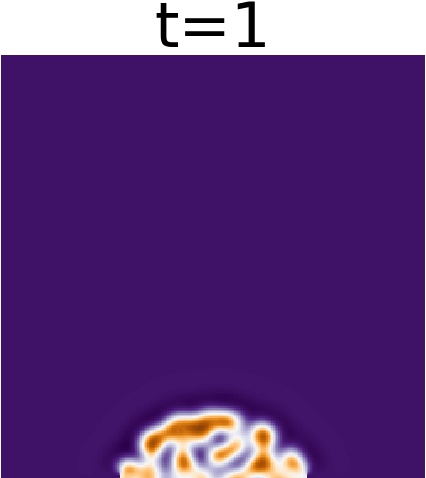}}
\subfigure{\includegraphics[height=0.16\linewidth]{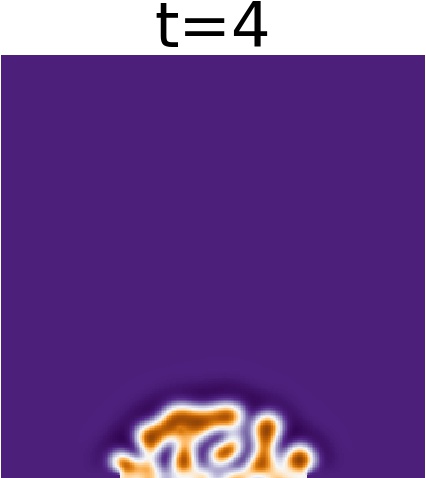}}
\subfigure{\includegraphics[height=0.16\linewidth]{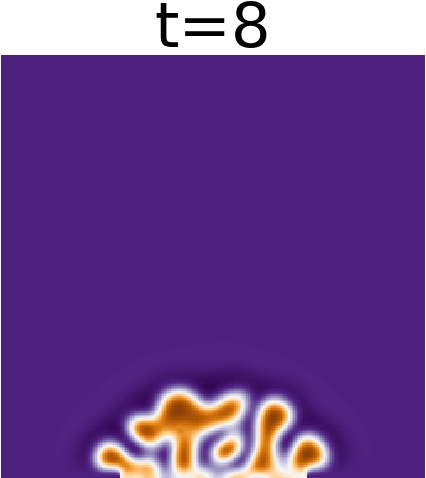}}
\subfigure{\includegraphics[height=0.16\linewidth]{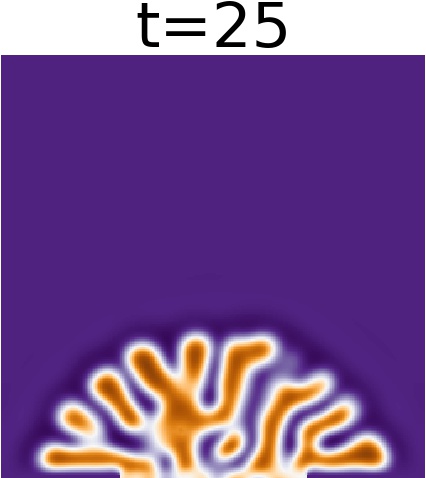}}
\subfigure{\includegraphics[height=0.16\linewidth]{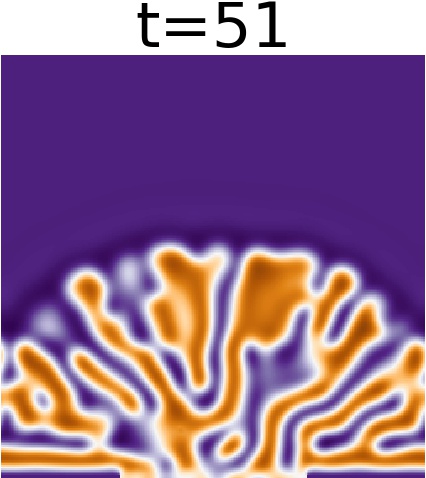}}
\subfigure{\includegraphics[height=0.16\linewidth]{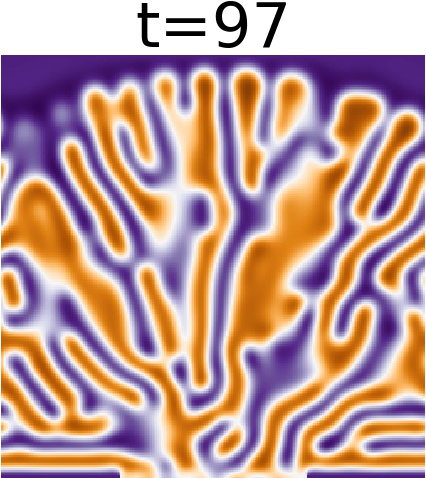}}
\subfigure{\includegraphics[height=0.14\linewidth]{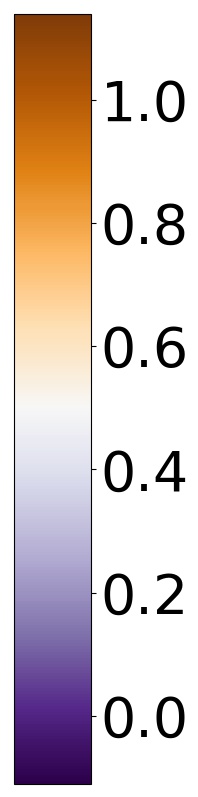}}
\caption{\label{fig:sim_seq}
Time sequence of a vertical cut of the field $f$ for a simulation starting with a noisy half sphere at the bottom of the simulation domain. 
Boundary conditions are periodic at the lateral boundaries, and $f=const$ at the top and bottom boundaries. 
The time is normalised by the growth rate of the most unstable mode in Fig. S1.} 
\end{figure}
\begin{figure}
\centering
\subfigure[t=38]{\includegraphics[height=0.4\linewidth]{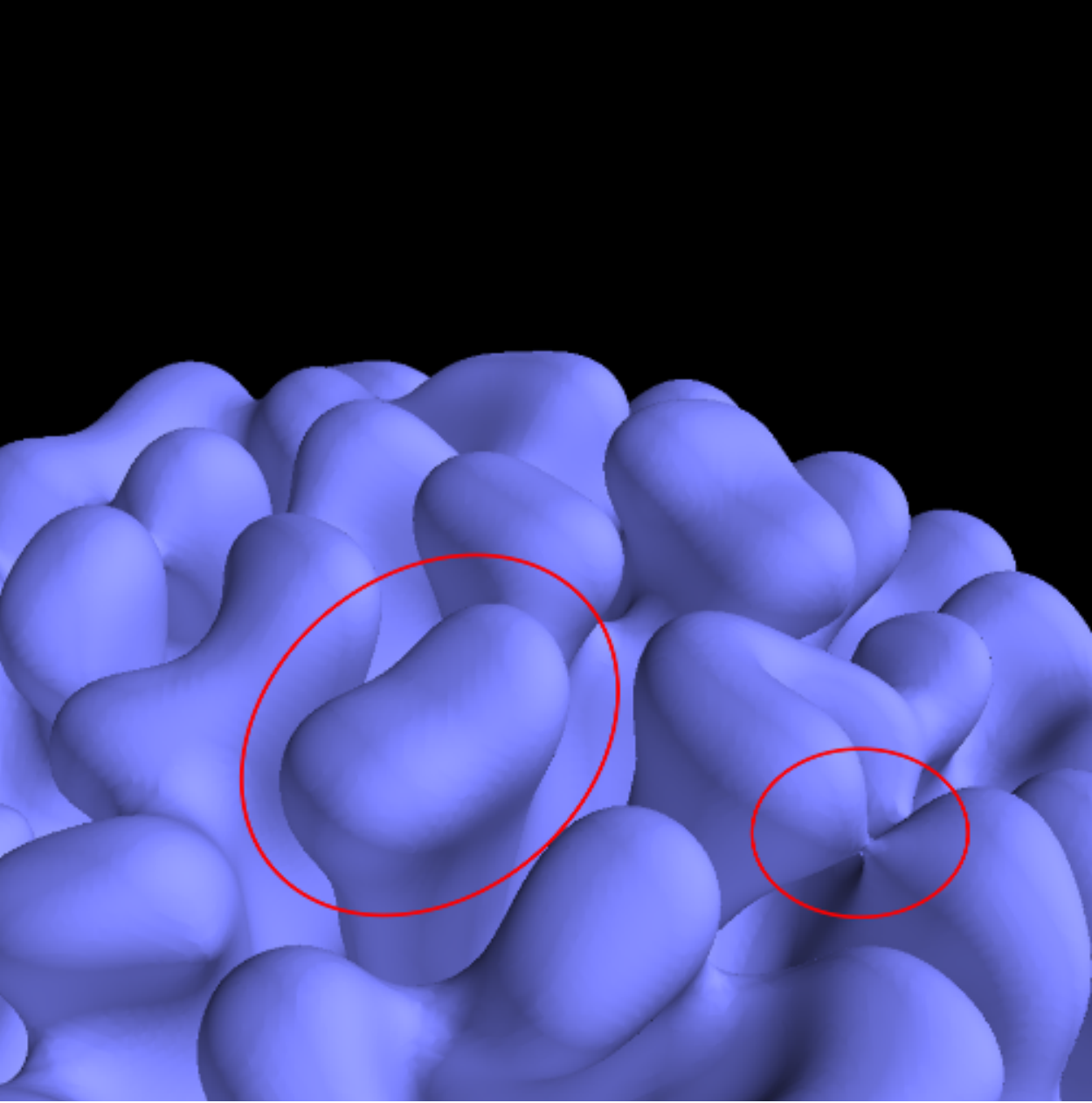}}
\subfigure[t=40.1]{\includegraphics[height=0.4\linewidth]{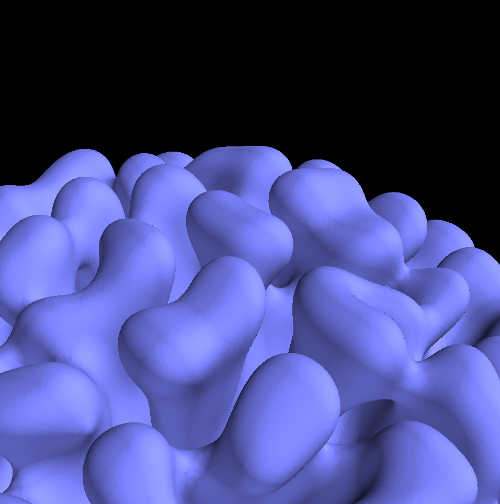}}

\subfigure[t=42.2]{\includegraphics[height=0.4\linewidth]{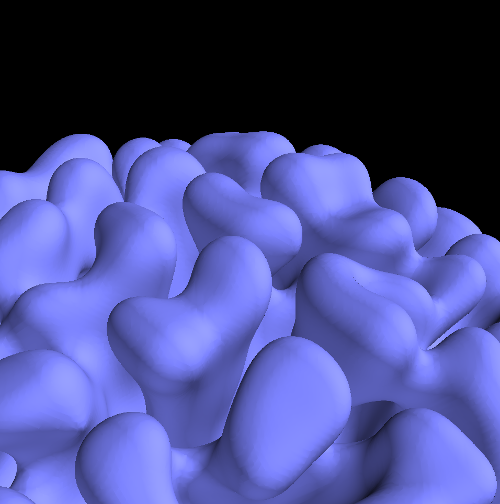}}
\subfigure[t=43.3]{\includegraphics[height=0.4\linewidth]{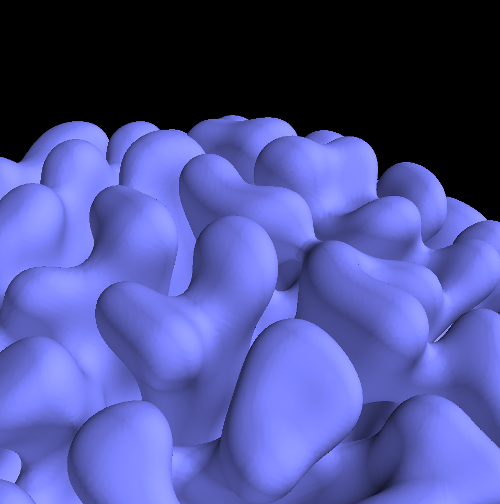}}
\caption{\label{fig:merge_and_branch}
Evolution of the $f=0.5$ iso-surface over a period of 5 time scales: one recognises at least one branching episode and one merging episode (red circles). 
The simulation is the same as the one described in Fig. \ref{fig:sim_seq}.
}
\end{figure}
As a first result we report that our model reproduces a growth process. 
In Fig. \ref{fig:sim_seq} we show the evolution of the field $f$ for a simulation initiated with a nest-like seed (see previous section) with the shape of a half-sphere at the bottom of an empty domain ($f=0$).
One observes that the initial seed grows, and progressively invades the empty space at a rate that is comparable with that of the most unstable mode in Fig. S1.
Even more interestingly we notice that far from the interface between the empty space and the growing nest the field $f$ is essentially stationary (for example there is little rearrangement of the pre-existing nest). 
This indicates that equation (\ref{eq:growth_0}) describes an accretion phenomenon happening at the interface between the growing walls and the empty space, even if we did not implement any freezing or aging effect and the field $f$ could, in principle, change everywhere and at any time.
Finally we remark that the growing interface frequently branches and merges as it is shown in Fig. \ref{fig:merge_and_branch}.
Coherently with the linear analysis, merging and branching episodes cover a few time scales.

Secondly we observe that in the growing pattern of Fig. \ref{fig:sim_seq} a characteristic length-scale appears which is the typical distance between two walls.
We report that a similar result was obtained for all the initial conditions and boundary conditions we considered.
In Fig. \ref{fig:sim_scale_length} (top) we show a two-dimensional slice of the $f$ field for two different simulations (a,b) where (a) has triply periodic boundary conditions and was initiated with white noise everywhere, while (b) is the one already presented in Fig. \ref{fig:sim_seq}. 
Both simulations were considered at large time, namely at $t=64$ time scales for simulation (a) and $t=93$ time scales for simulation (b).
To track the existence of a characteristic length-scale in our objects we perform a three-dimensional fast Fourier transform (FFT)  which is shown in Fig. S2.
The dominant frequency of the FFT was then obtained and superimposed (squares) to the patterns of Fig. \ref{fig:sim_scale_length} as a comparison. 
We remark that the dominant length indicated by the FFT is consistent with the observed patterns.
Finally we note that the emerging lengths are fairly consistent with the most unstable length indicated by linear analysis as shown in Fig. S1.
Nonetheless the observed patterns significantly differ from the planar solutions we consider in section \ref{subsec:linear}, and appears to be modulated in multiple directions. 
To quantify this feature we compute the auto-correlation functions $C_f(x)$, $C_f(y)$, $C_f(z)$   in the three Cartesian directions (symbols) and the function $C_f(|\boldsymbol{x}|)$ averaged over all directions (solid line), with all the distances rescaled by the typical length given by the FFT. 
The results are shown at the bottom of Fig. \ref{fig:sim_scale_length}.
First, one observes that in all the diagrams $C_f(|\boldsymbol{x}|)$ shows a first maximum around $d=1$ which is consistent with FFT analysis. Then we remark that in simulation (a) all the directional correlation functions present the same peak as $C_f(|\boldsymbol{x}|)$ which indicates that the observed pattern is highly isotropic, while in simulation (b) there is no peak in the vertical correlation function $C_f(z)$, which seems to indicate that there is low modulation in the vertical direction.
Indeed, all the simulations we performed ultimately show a similar sheet-like anisotropy. 
This happens after a very slow process of rearrangement which usually takes tens or even hundreds of time scales after the pattern has invaded the entire domain.
Our interpretation is that this particular pattern is always selected by nonlinear interactions or secondary instabilities similar to those observed in elastic plates \citep{Mahadevan2005,Audoly2008}, but different initial conditions can enhance or not this effect, determining at which time this pattern takes over.

\section{Comparison with real nests}\label{sec:comparison}
\begin{figure}
\centering
\subfigure[a][]{
\begin{overpic}[width=0.23\linewidth]{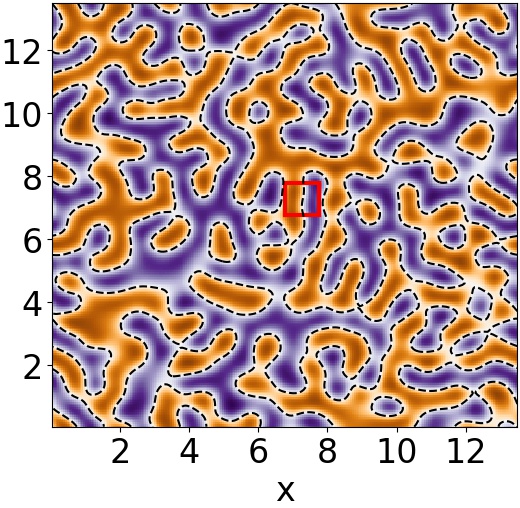}
\put(-8,55){\small $z$}
\end{overpic}
}
\hfill
\subfigure[b][]{\includegraphics[width=0.23\linewidth]{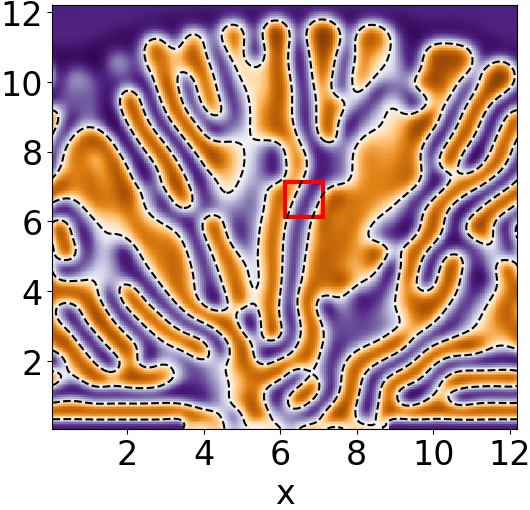}}
\hfill
\subfigure[c][]{\includegraphics[width=0.23\linewidth]{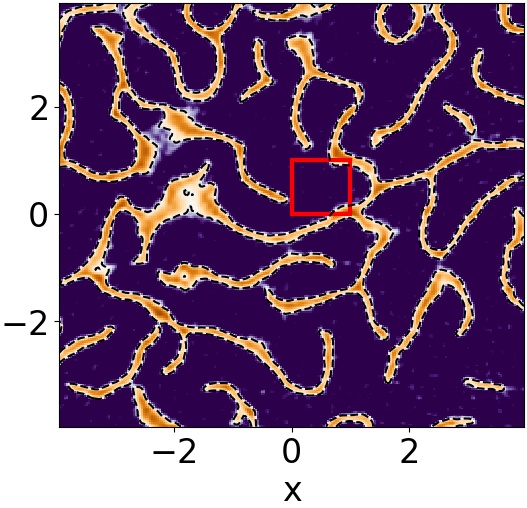}}
\hfill
\subfigure[d][]{\includegraphics[width=0.23\linewidth]{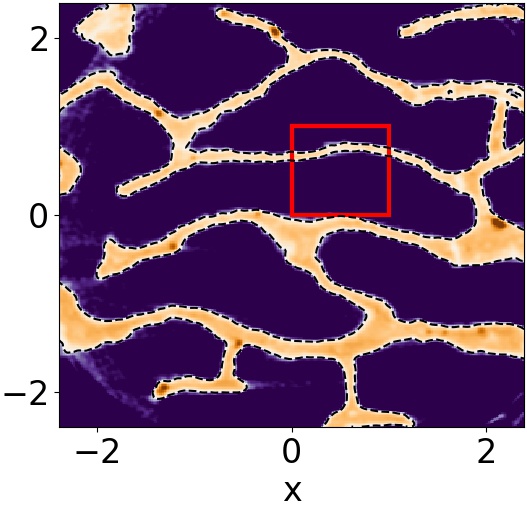}}

\subfigure{
\begin{overpic}[width=0.23\linewidth]{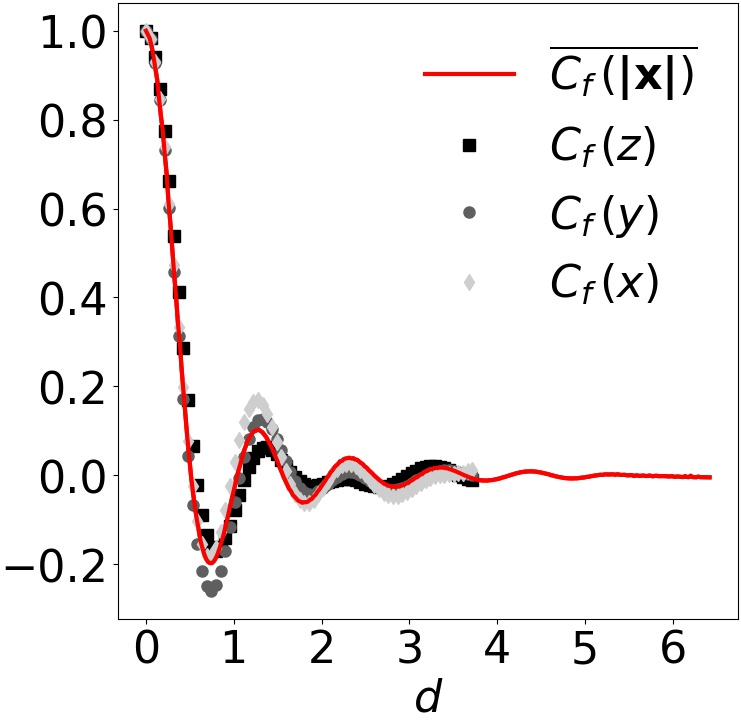}
\put(-8,55){\tiny $C_f$}
\end{overpic}
}
\hfill
\subfigure{\includegraphics[width=0.23\linewidth]{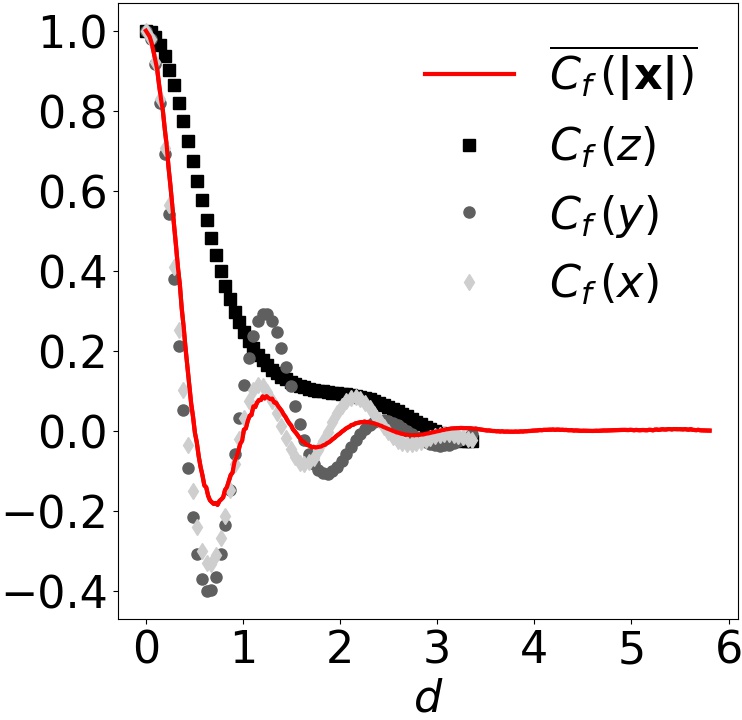}}
\hfill
\subfigure{\includegraphics[width=0.23\linewidth]{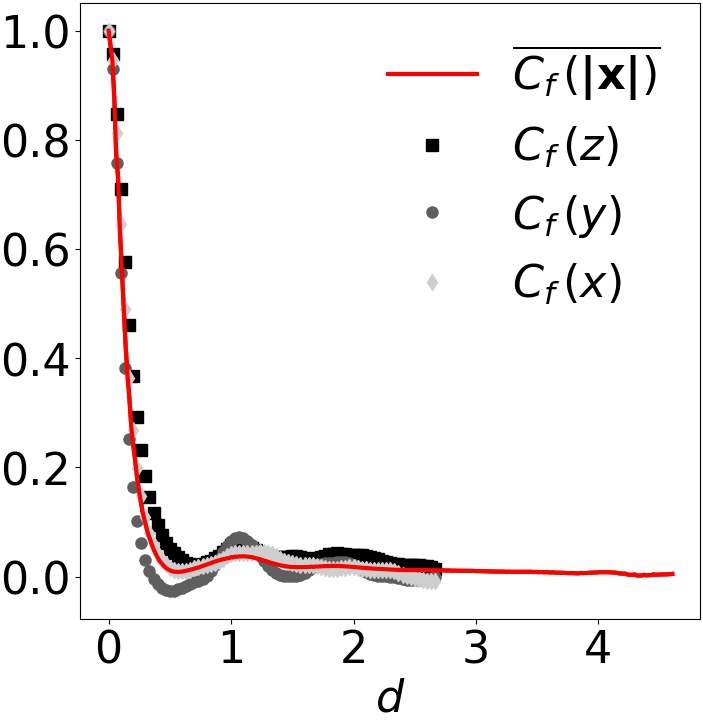}}
\hfill
\subfigure{\includegraphics[width=0.23\linewidth]{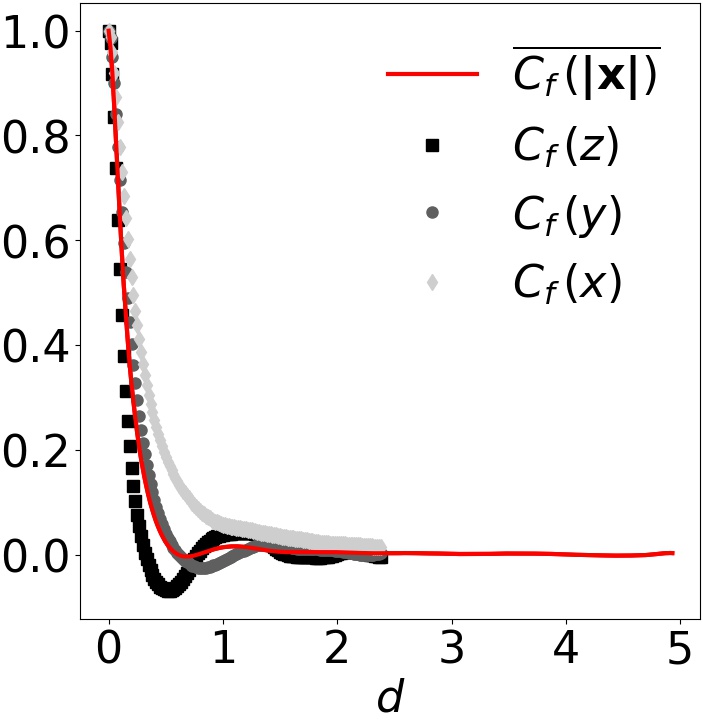}}
\caption{\label{fig:sim_scale_length}
Top: Cross section of the nest volume for two simulations (a,b) and two ct-scans of real samples (c,d).
The walls are in orange and the empty space is in blue. The black square indicates the peak of the 3D fast Fourier transform over the whole volume.
Bottom: Auto-correlation functions $C_f$, each one corresponding to the nest above it.
Symbols correspond to the value of $C_f$ when computed only in one Cartesian direction, while the solid lines correspond to the average value of $C_f$ over all possible directions.   
}
\end{figure}

\begin{figure}
\centering
\subfigure{\includegraphics[height=0.32\linewidth]{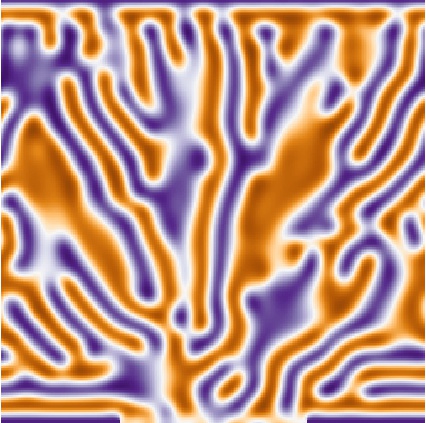}}
\hfill
\subfigure{\includegraphics[height=0.32\linewidth]{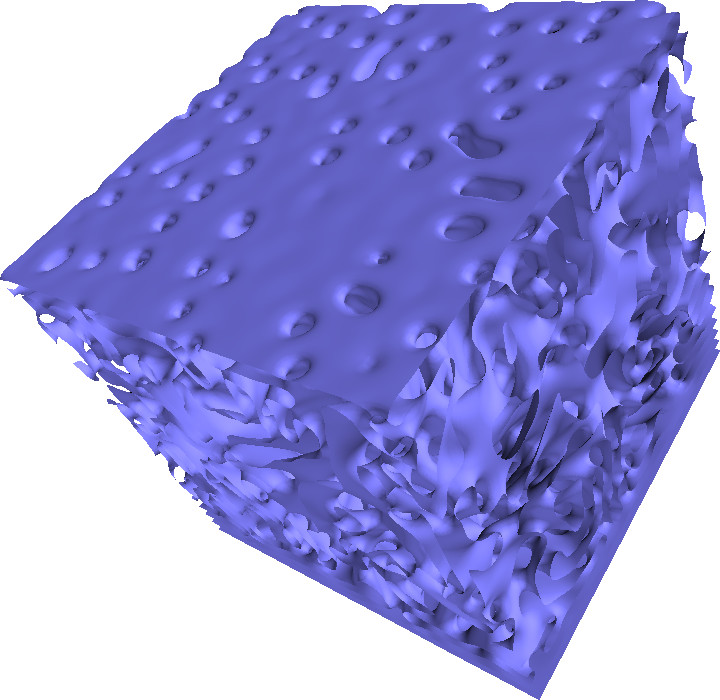}}
\hfill
\subfigure{\includegraphics[height=0.32\linewidth]{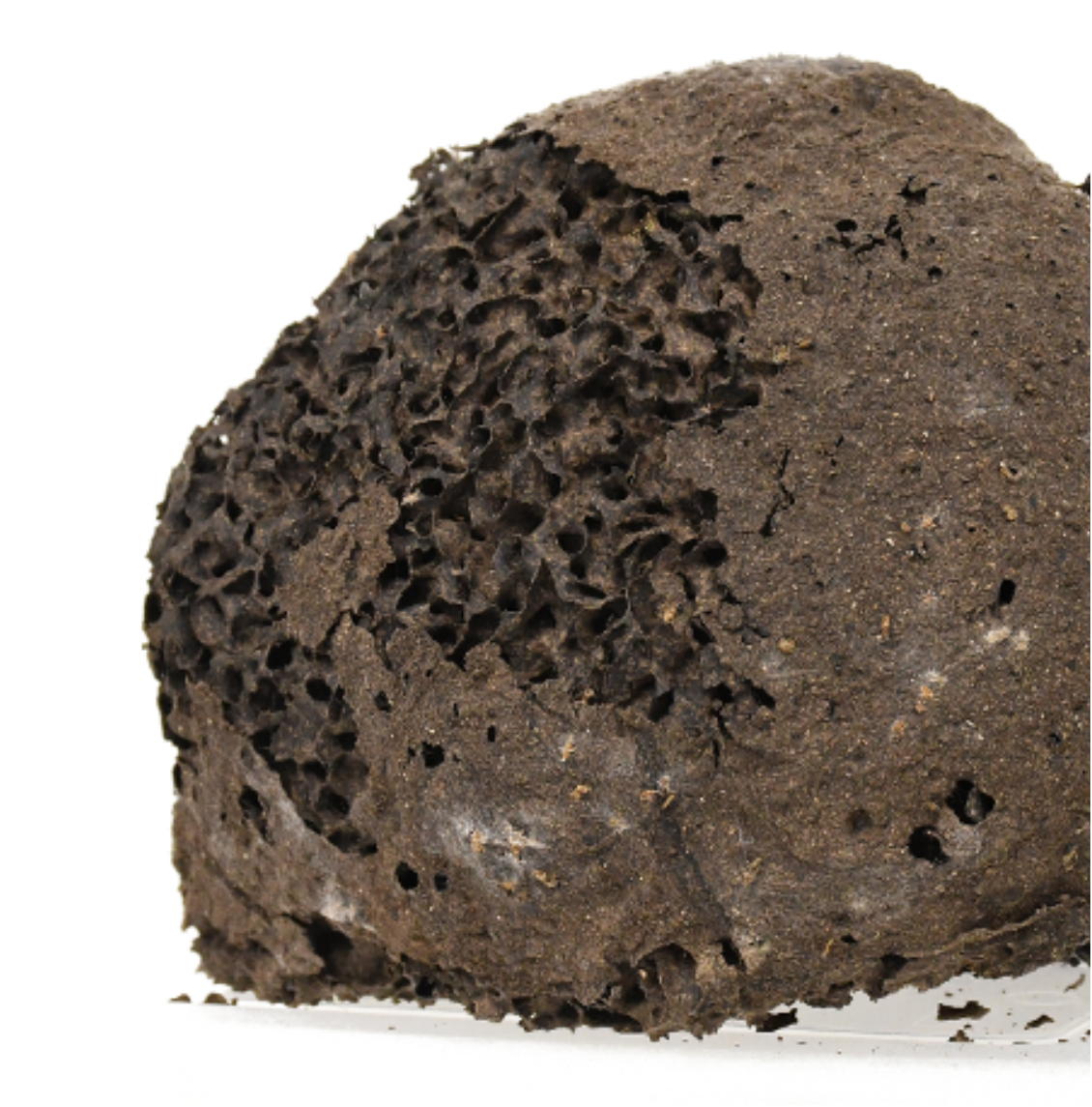}}
\caption{\label{fig:BC_effect}
Effect of a $f=0$ boundary on the nest growth. Left and center panel correspond to a 2D vertical cut and a 3D rendering of simulation (b) at $t=128$.
The top face of the simulation domain was maintained constant at $f=0$.
One observes, that approaching the forbidden boundary $f=0$ the simulated nest closes itself forming a roof which resembles the thin external layer of \textit{Nasutitermes} nests (right).
} 
\end{figure}

We collected nest fragments of two different species of \textit{Nasutitermes}, namely \textit{N.walkeri} from the Sydney area (Australia) and \textit{N.ephratae} from Guyana; two of these are shown in Fig. \ref{fig:nasuti_nests}. 
The walls of these nests are sometimes infra-millimetric thus a correct reconstruction of their structure was possible only with the use of a micro-CT scanner whose resolution was set to $\unit[0.1]{mm}$.
A typical scan consists of a stack of images, where higher intensity corresponds to the walls and lower intensity to the empty space, analogously to the field $f$ that we consider in our model and simulations.
We repeated our scale analysis on two ct-scans of fragments belonging to real nests of the species \textit{N. walkeri} (c) and \textit{N. ephratae} (d) which are reported in Fig. \ref{fig:sim_scale_length}. As for simulated nests, the FFT of real nests indicates a dominant length-scale which is consistent with the observed distance between walls (see Fig. S2).
Looking at the auto-correlation functions one sees that fragment (c) is poorly modulated in the horizontal direction, which is consistent with the observed pattern, while fragment (d) appears fully isotropic. 
Note that for these fragments vertical and horizontal directions were completely arbitrary.
Indeed different degrees of isotropy or different local structures can be encountered in different species and even within the same nest at different depths, as in the case of \textit{N. ephratae} \citep{Thorne1980}.
Interestingly \cite{Jones1979} reported that in \textit{N. costalis}, if the nest is cut open along a plane parallel to the growth direction one observes a tall columnar structure whose description highly resembles our simulation (b). 
As a summary our model appears complex enough to reproduce the variety of structures encountered in real nests as a response to different initial conditions, which suggests that sensitivity to initial conditions also matters in the real growth process.
As a second comparison we report that whenever the growing interface approaches a boundary where $f$ is set to $0$, the growing tips tend to connect forming a sort of roof scattered with holes, as it is shown in Fig. \ref{fig:BC_effect}. 
Waiting even longer the holes tend to be filled and the connections between the roof and the rest of the structure are progressively eroded.
Very interestingly in all the species of arboreal \textit{Nasutitermes}, nests are covered with a uniform thin crust which in species like \textit{N. ephratae} happens also to be poorly connected to the bulk structure \citep{Jones1977,Thorne1980}. 

In the present literature there is no comprehension of what could control the activation/inhibition of a construction stage, which is usually extremely rapid compared to the lifetime of a colony \citep{Jones1979}. 
Establishing a direct correspondence between real nests and simulations on this question may be premature, however we consider extremely encouraging that such a simple model incorporates a boundary sensitivity that can mimic the real process of finishing up the nest, which likely involves several other external (weather conditions, material availability, etc) ad internal (size and health of the colony) variables, that have been completely ignored at this stage.
Moreover, we could observe in the field that whenever the external layer of a thriving colony is disrupted, termites rapidly construct a patch that seals the internal structure from the exterior. 
This suggests that boundary conditions could mimic the maintaining of a non expanding stage. 

\begin{figure}
\centering

\subfigure{
\begin{overpic}[width=0.45\linewidth]{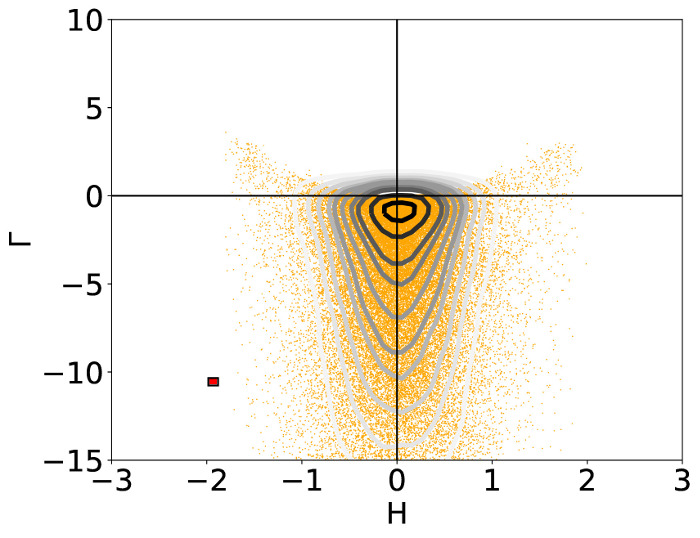}
\put(85,65){(a)}
\end{overpic}
}
\hfill
\subfigure{
\begin{overpic}[width=0.45\linewidth]{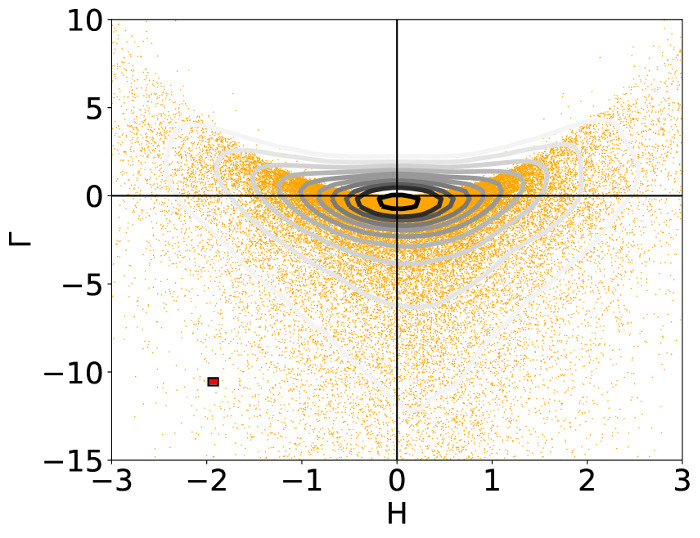}
\put(85,65){(b)}
\end{overpic}
}

\subfigure{
\begin{overpic}[width=0.45\linewidth]{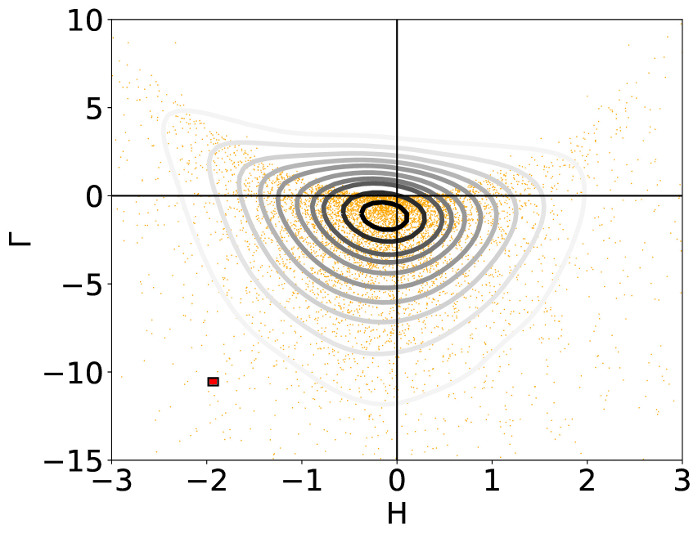}
\put(85,65){(c)}
\end{overpic}
}
\hfill
\subfigure{
\begin{overpic}[width=0.45\linewidth]{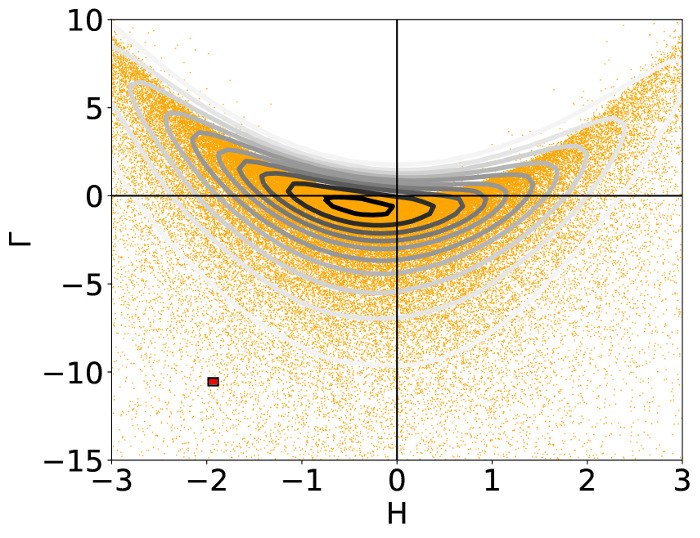}
\put(85,65){(d)}
\end{overpic}
}

\subfigure{
\begin{overpic}[width=0.45\linewidth]{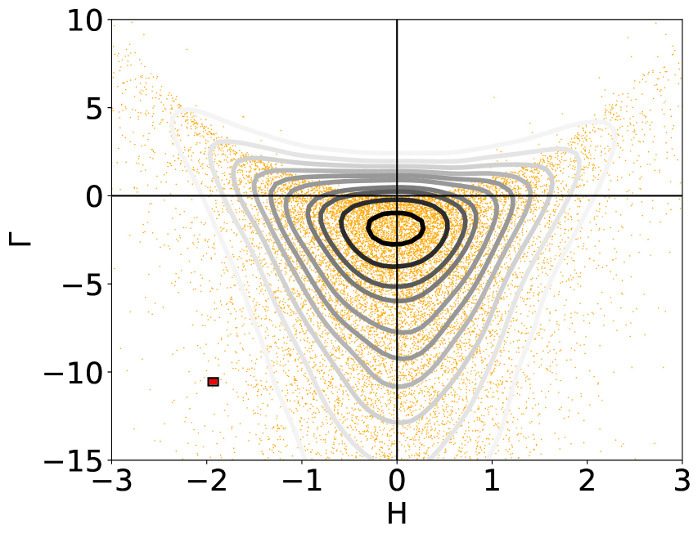}
\put(85,65){(e)}
\end{overpic}
}
\hfill
\subfigure{
\begin{overpic}[width=0.45\linewidth]{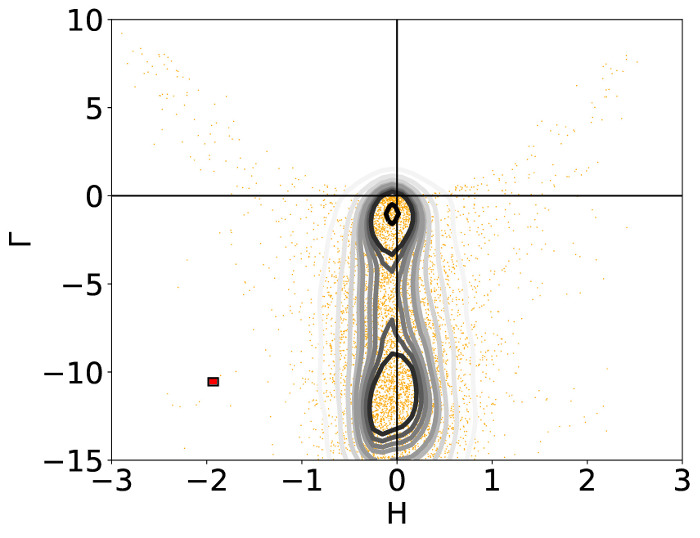}
\put(85,65){(f)}
\end{overpic}
}
\caption{\label{fig:HG_kde}Kernel-density-estimation KDE of the frequency of surface mesh elements in the bidimensional space mean curvature ($H$) - Gaussian curvature ($\Gamma$). 
Each contour corresponds to the successive 10 percentile of points from more frequent area (black) to less frequent area (light gray).
Dots represent surface elements (one over four dots is shown). The small square indicates the bin size for KDE.
}
\end{figure}

At this point we have already established a strong correspondence between our model and real nests, which are: the very convoluted structure (frequent merging and branching), the existence of a typical scale, 
the sensitivity to external constraints both in space (boundary conditions) and time (initial conditions). 
However all these are mostly qualitative comparisons.
Below we define a protocol to make our comparison quantitative.
The difficulty of this task is due to the fact that our objects are fundamentally disordered so that a statistical approach is the only possible route. 
The choice of the relevant quantity to compare may also be discussed, but coherently with our observations of real nests and with our growth model, the local curvature is the best candidate.
The local curvature of a 3D surface in one point $P$ can be entirely defined by two principal curvatures $k_1$ and $k_2$ which are defined as the minimum and the maximum among the curvatures $k$ of all the normal sections, i.e. the planar curves given by the intersection between the surface and a normal plane. Given the osculating circle to such a curve, the value of $k$ is the reciprocal of its radius while the sign is positive or negative depending on whether the curve turns in the same or the opposite sense of the local normal vector $\boldsymbol{\hat{n}}$. 
Equivalently one can define the mean curvature $H$ (that we already encountered in sec. \ref{sec:model})
and the Gaussian curvature $\Gamma$:
\begin{equation}\label{eq:HG_def}
H=(k_1+k_2)/2 \;\;\;\;\;\;\;\; \Gamma=k_1k_2    
\end{equation}
At this stage one may simply keep in mind that the sign of $\Gamma$ tell us if the surface is more similar to a sphere ($\Gamma>0$), a cylinder ($\Gamma=0$), or a saddle ($\Gamma<0$), while the sign of $H$ tells us if we are at the exterior ($H>0$), or the interior ($H<0$), of our spherical, cylindrical, or saddle-like surface. 
Finally $H = 0$ characterize peculiar saddle-like surfaces which are  known as \textit{minimal surfaces} where no interior or exterior can be defined, the two sides being indistinguishable.
To define our statistical sample we take advantage of the discrete representation of our objects. 
As we defined them, both simulations and real nests are scalar fields defined on a grid of voxels, from where we extract an iso-surface in the form of three-dimensional triangular meshes obtained via the Lewiner Marching Cube \citep{Lewiner2003} algorithm implemented in \textit{Python}.
Our sample space will then be constituted by the set of all surface elements, to which we associate two independent attributes which are the mean curvature $H$ and the Gaussian curvature $\Gamma$.
These are obtained as a function of the mesh vertices coordinates as described in section S.II; 
importantly these coordinates are priorily normalised by the typical length given by FFT in order to make the comparisons consistent. 
In Fig. S3 we report a histogram of the frequency of our two observables $H$ and $\Gamma$ for a collection of CT-scans of real nest fragments, simulations and synthetic data. 
The distribution of $H$ is always peaked and quite symmetric around $H=0$ while the values of $\Gamma$ are more frequent at $\Gamma<0$ which indicates that all our surfaces share the abundance of saddle-shaped region.
Nonetheless a full characterization of our objects demand to consider $H$ and $\Gamma$ at once.

In Fig. \ref{fig:HG_kde} we have reported the cumulative Gaussian kernel density estimation KDE 
(see equation S2) 
of the surface elements in the space $(H,\Gamma)$.
The regions between two contours indicate where ten percent of the total counts occur, from the most frequent area, in black, to the most rare area, in light gray.
In the top and central rows we have reported the two simulations (a-b) and two fragments of real nests (c-d) already analysed in Fig. \ref{fig:sim_scale_length}. At the bottom we considered two synthetic benchmark surfaces which are a random surface (e) obtained from a volume of white noise which has been band-pass filtered around a certain length, and a gyroid (f) which is a minimal surface, i.e $H=0$ everywhere.
One recognises a strong resemblance in the distribution between simulation (b) and nest (d). Compared to all the other diagrams one observes that  the curvature distribution is pushed  toward the center $H=\Gamma=0$ and its shape is squeezed in the direction of the $\Gamma$ axis.
This distinction is coherent with the one we have made in Fig. \ref{fig:sim_scale_length} in term of isotropy/anisotropy, since in simulation (b) and nest (d) we observe an higher proportion of quasi-flat regions which is consistent with higher anisotropy.
Conversely simulation (a) and nest (c) are more isotropic and share a broader distribution along the $\Gamma$ axis which indicates an higher abundance of saddle-shaped regions and is closer to the curvature distribution of a gyroid (f).
One also remarks that in nest (c) curvature distribution is more spread along the $H$ axis, and it is extremely similar to the curvature distribution of the random surface (e).
These results suggest that the construction process is intrinsically noisy and that probabilistic instead of deterministic rules should control the building behaviour \citep{Invernizzi2019,Camazine2001}. 
Interestingly our deterministic model manages to capture this feature when we let the system to invade the empty space from a growing boundary as in simulation (b) but the noise is partially removed if we assign a random initial condition to all the volume and let the system evolve for long time as in simulation (a).

\section{Discussion}\label{sec:discussion}
Construction, as performed by humans or other animals, usually involves the interaction between a building agent and some building material under the guidance of innate or learned rules of building. 
In contrast, growth - as it is known in physics or biology - is an autonomous process where a substrate grows by itself, depending on its interaction with the external environment.
Arboreal nests built by \textit{Nasutitermes} termites lie in the middle between these two categories because they originate from the cooperation of thousands of individual workers but the interactions between them is largely guided by the shape of the substrate itself which makes the nest construction a self-organized process.  
This justifies our approach of formulating a model where the agents' behaviour is incorporated in a classical accretion process where the nest growth is focused in the region of high curvature. In spite of its simplicity, our model is sufficient to reproduce some of the global and local properties of arboreal \textit{Nasutitermes} nests, namely the intricate structure, the existence of a typical length-scale, the sensitivity to boundary and initial conditions, and the abundance of saddle-shaped structure of zero mean curvature.
We model the nest as a continuous phase endowed with a scalar field $f$ whose intensity delimits the nest walls or cavities according to a threshold value. The nest growth can then be described by a single nonlinear equation that includes a growing term proportional to local curvature and a dissipative term which is sensitive to small spatial scales, their relative importance being fixed by one adjustable parameter $d$. 

First we report that when $d$ is large enough our minimal equation successfully reproduces a growth process where the growing substrate progressively invades the available space, through a rich dynamics of branching and merging which produces intricate structures similar to \textit{Nasutitermes} nests. 
We stress here that while branching is commonly observed in many gradient driven growth phenomena both in inert \citep{Pelce2000} and biological matter \citep{Kaandorp1999,Clement2012} merging is impossible in such simple models \citep{Lajeunesse2000}. Only recently when the gradient is reintroduced on both sides some merging was observed \citep{Budek2017}.
With regard to our model, we report that merging occurs only in 3D, while in 2D the symmetry $f \rightarrow (1-f)$ must be broken in order to observe it.
We also report that, far from the front of freshly grown substrate, rearrangement of the interface is negligible, although it is permitted.
This is consistent with the empirical observations by \cite{Jones1979,Thorne1996} which report that in \textit{Nasutitermes}, nest walls are barely rearranged after the initial construction.

All the simulated patterns clearly show the emergence of a characteristic length which corresponds to the one predicted by linear analysis and scales as $1/\sqrt{d}$. This feature constitutes a strong similarity with nest samples collected in the field which show a characteristic distance between neighboring walls. 
In real nests such a distance is comparable with termites body size, which suggest that termites may use their body as a template as recently proposed for ants \citep{Khuong2016}. 
However our model shows that there is no need to enforce the use of a template to observe the appearance of a typical distance.     
Interestingly at the linear order our problem is analogous to the buckling instability of a thin elastic plate under a symmetric in-plane compression of magnitude $d$, when $d$ is large enough the plate buckles and the compression energy is released in the form of bending energy of the plate. 
This problem was originally addressed more than one century ago but has received renewed attention in recent studies that showed how in this case the bending patterns usually differ from the planar waves predicted by linear analysis \citep{Audoly2008}. Instead richer patterns appear which are the combination of multiple planar waves oriented in different directions, because of secondary instabilities and nonlinear interactions. A direct comparison between our model and the thin plate problem is not possible because of the different dimensionality and initial conditions. 
Yet, it is interesting to notice how, similarly to the thin plates problem, in our simulations we never obtain the planar solution but rather observe truly three-dimensional patterns with a variable degree of isotropy depending on the initial conditions. As a general rule our equation tends to select structures which have a narrow wavelength in one direction and are slightly modulated in the two others.
Incidentally we observe that nest fragments from different termites species, or even within the same nest, also show a distinct variability in term of isotropy, thus we speculate that initial conditions should be relevant in the growth of real nests as well, but further investigations will be necessary in this regard. 
Generated patterns are also very sensitive to boundary conditions and in particular we observe that forbidding the nest expansion (e.g. imposing $f=0$) at one boundary induces the formation of a uniform layer that isolates the nest from the forbidden boundary. 
Interestingly all \textit{Nasutitermes} nests are also covered with an external thin layer which seals the nest from the outside world and it is constantly repaired whenever damaged by wetting/drying episodes or other external mechanical stresses, as it was observed both in the field and in the laboratory. 
A forbidden boundary does not necessarily correspond to the existence of a solid boundary but could correspond for instance to a threshold-level of pheromone emanating from the colony, as in the recent modelling study by \cite{Ocko2019}. In any case we find relevant that our simple construction model responds in a consistent way in term of expressed morphology.
Future empirical studies would be needed to identify the releaser stimuli that trigger the initiation and termination of nest expansion.

Both the real nests and the simulated patterns include many saddle-shaped regions where the mean curvature is zero. 
This is also shown by our quantitative comparison based on the curvature distribution, that maps our objects to a specific distribution in the two-dimensional space of mean ($H$) and Gaussian curvature ($\Gamma$). 
Both the simulated and scanned surfaces show large portions of zero mean curvature and non positive Gaussian curvature, which corroborates the hypothesis that curvature is the main driving mechanism in construction. 
In fact if growth happens in the region of high mean curvature, we expect that a stationary or quasi-stationary configuration should correspond to one where local mean curvature is low or null.
Consistently, whenever the structure is non trivial (e.g. different from a plane), highly connected and isotropic, we find that Gaussian curvature is negative $\Gamma<0$. 
Alternatively we observe both in simulations and real nests the existence of sheet-like structures of the type ($H\sim 0$, $\Gamma\lesssim 0$) which corresponds to the emergence of a clear anisotropy.
We also remark that curvature distribution of isotropic nests resembles that of a random synthetic surface with a typical length-scale, which confirms the absence of a global predetermined architecture. This is consistent with a construction process driven primarily by local information, as we implemented in our model. 

While nest structure is primarily random on the local scale, we already observed that nest fragments can have different degrees of internal anisotropy and distinctive features such as the external envelope. 
Due to the relatively small size of our samples, we do not have a detailed characterisation of nest-topology and connectivity on the global scale and the question remains open, if the observed topology and connectivity are an emerging property of a same local construction mechanism - as in the present model - or they must be enforced through additional global rules.
On another side, we plan to deeper investigate the reasons that trigger the beginning or the end of a construction stage.
Previous literature is all but conclusive in this regard but we expect that activation and inhibition should be related to the colony size, which can be possibly translated in the abundance of some auxiliary field, like relative humidity, CO$_2$ or a pheromone.
To this aim we will couple our model with some auxiliary field that is constrained to diffuse or convect in the nest galleries and explore whether we can observe significant time modulation in the nest expansion.

In the general context of accretion processes we believe that our equation may be adapted to other phenomena where a growth rate sensitive to curvature and a local smoothing process coexist, which is encouraged by the simplicity of the equation and the richness of the expressed patterns. 
While our model was primarily formulated to describe the morphogenesis of arboreal termite nests, saddle-shaped structures with a characteristic scale are also observed in other biological systems, ranging from the nests of other insects not directly related to termites (e.g. \textit{Lasius fuliginosus}), to the micro-structure of butterfly wings \citep{michielsen2008gyroid}. It is possible that our model could provide insight also on the formation of these other structures.
The scientific literature on social insects has often focused on agent-based models as a tool to describe the behaviour of individual insects and the emerging organisation of the colony. We are confident that simple ``stigmergic'' models like the one developed here have a great potential to explain not only how structures are built, but also how transitions between different nest plans can be triggered by small changes in the building parameters or environmental conditions. In relation to real nests, such changes could correspond to transitions between nest plans that emerge across phylogenies and could also reflect adaptive changes in response to different habitats.

\bibliographystyle{abbrvnat}
\bibliography{nasute}
\end{document}


\maketitle
\section{Linear and scale analysis}
\begin{figure}[b!] 
\centering
\includegraphics[width=.7\linewidth]{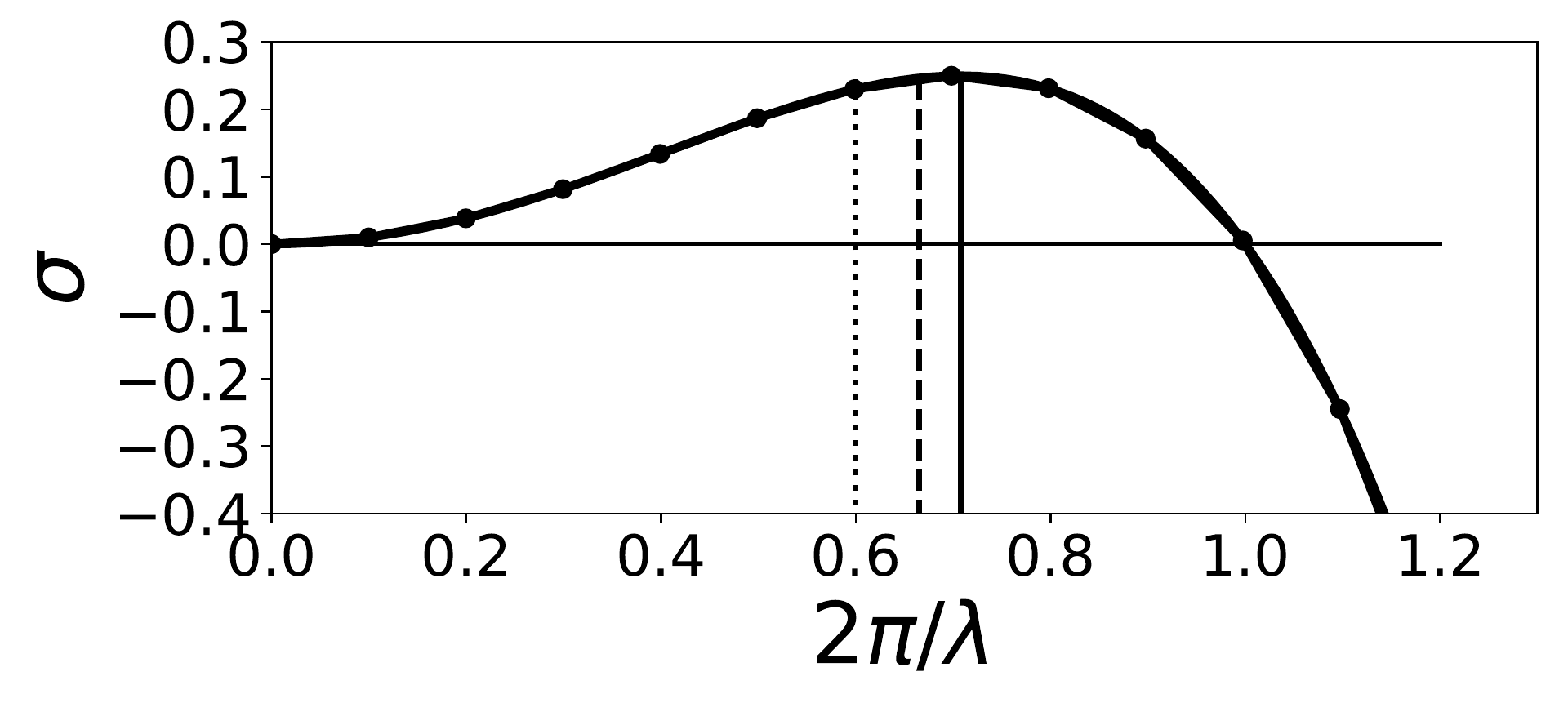}
\caption{\label{sfig:disp_relation} Dispersion relation \ref{eq:disp_relation} for small-amplitude perturbations, here $\tilde{f}=0.25$ and $d=4$. The solid line indicates the most unstable length scale while dashed and dotted lines 
indicate the typical length that appears in simulations (a) and (b) reported in figure 
4.
Black dots refer to lengths that can fit in our simulation box.}
\end{figure}
\begin{figure}[t!] 
\centering

\subfigure{\includegraphics[width=0.24\linewidth]{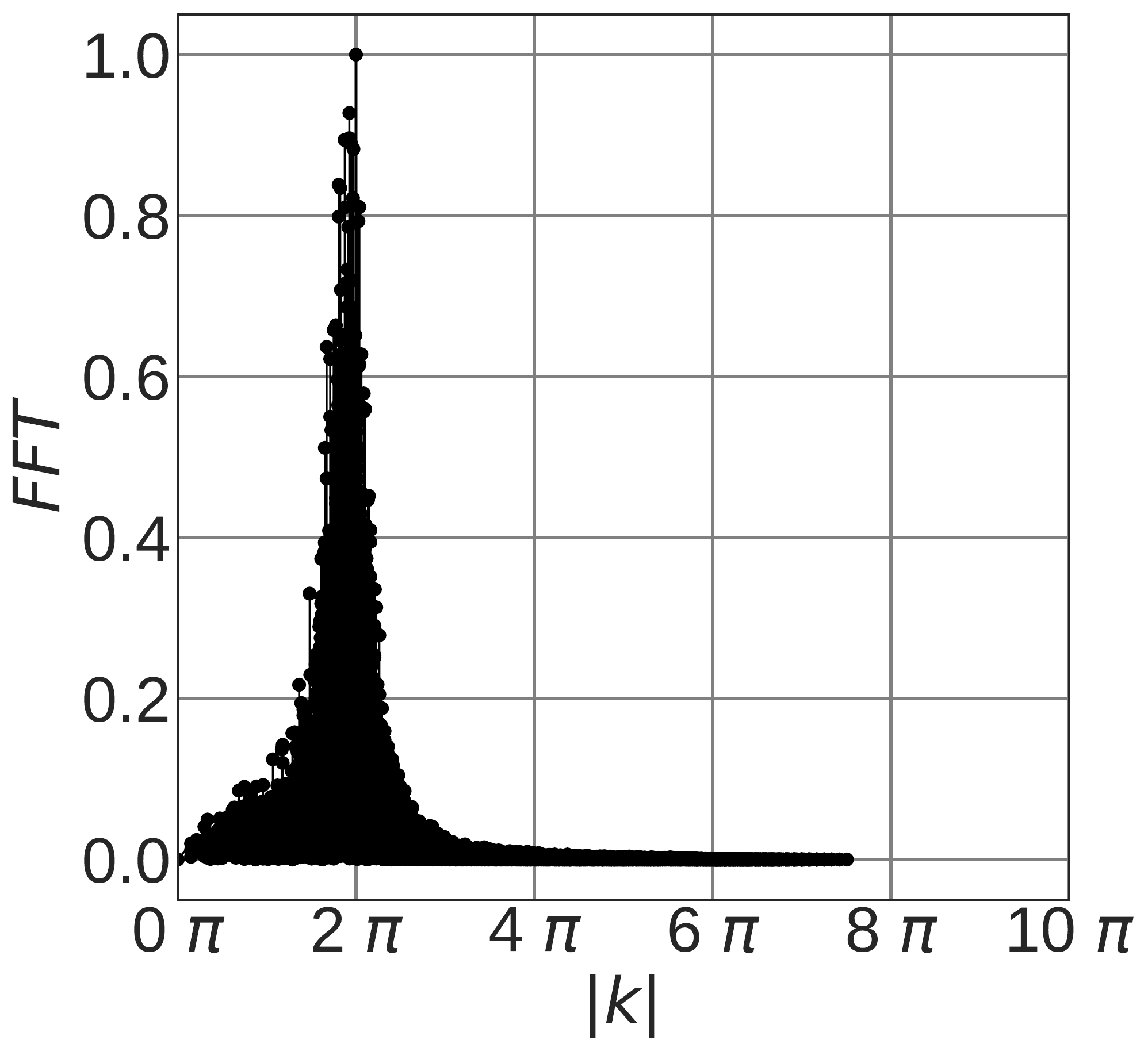}}
\subfigure{\includegraphics[width=0.24\linewidth]{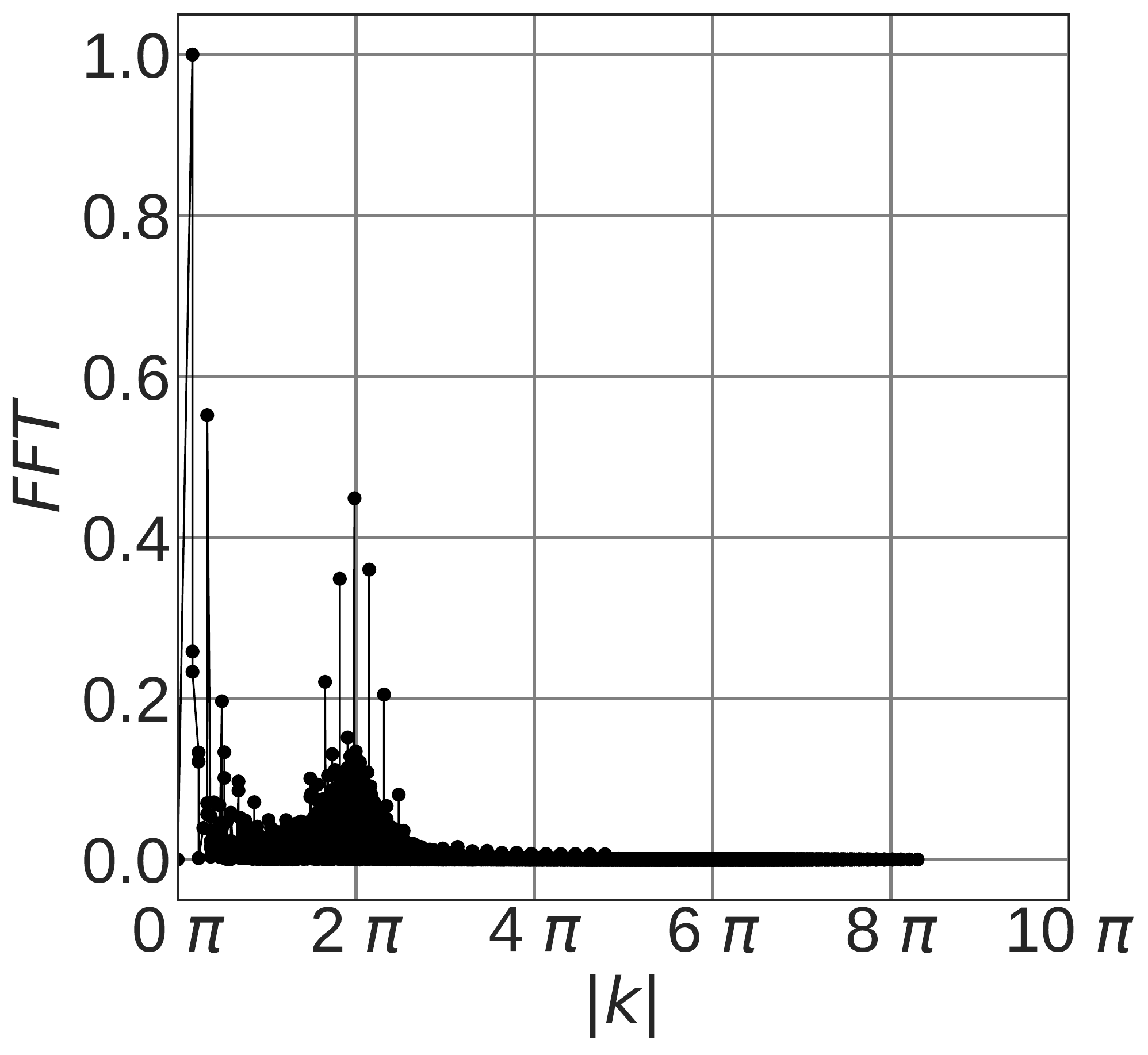}}
\subfigure{\includegraphics[width=0.24\linewidth]{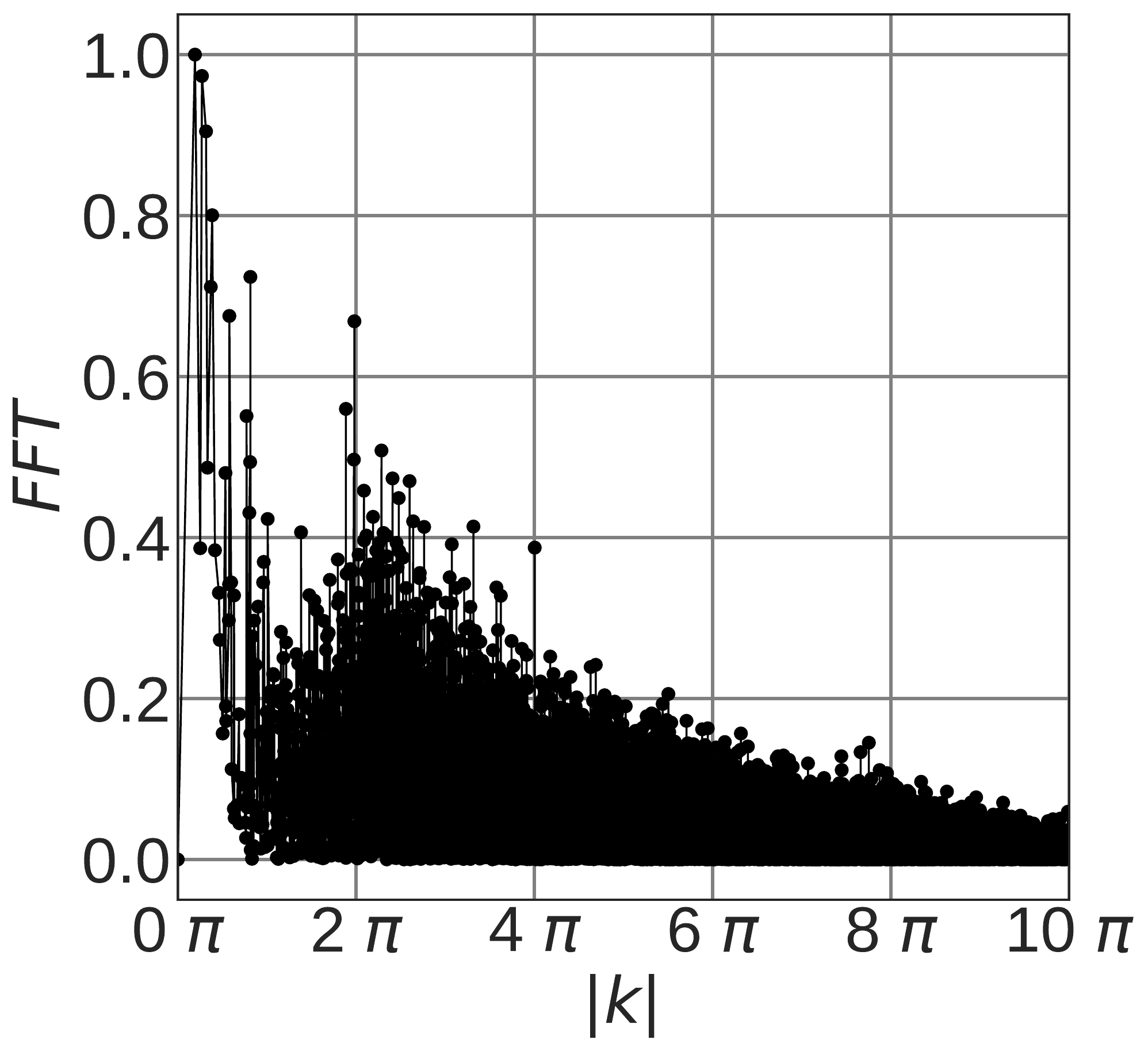}}
\subfigure{\includegraphics[width=0.24\linewidth]{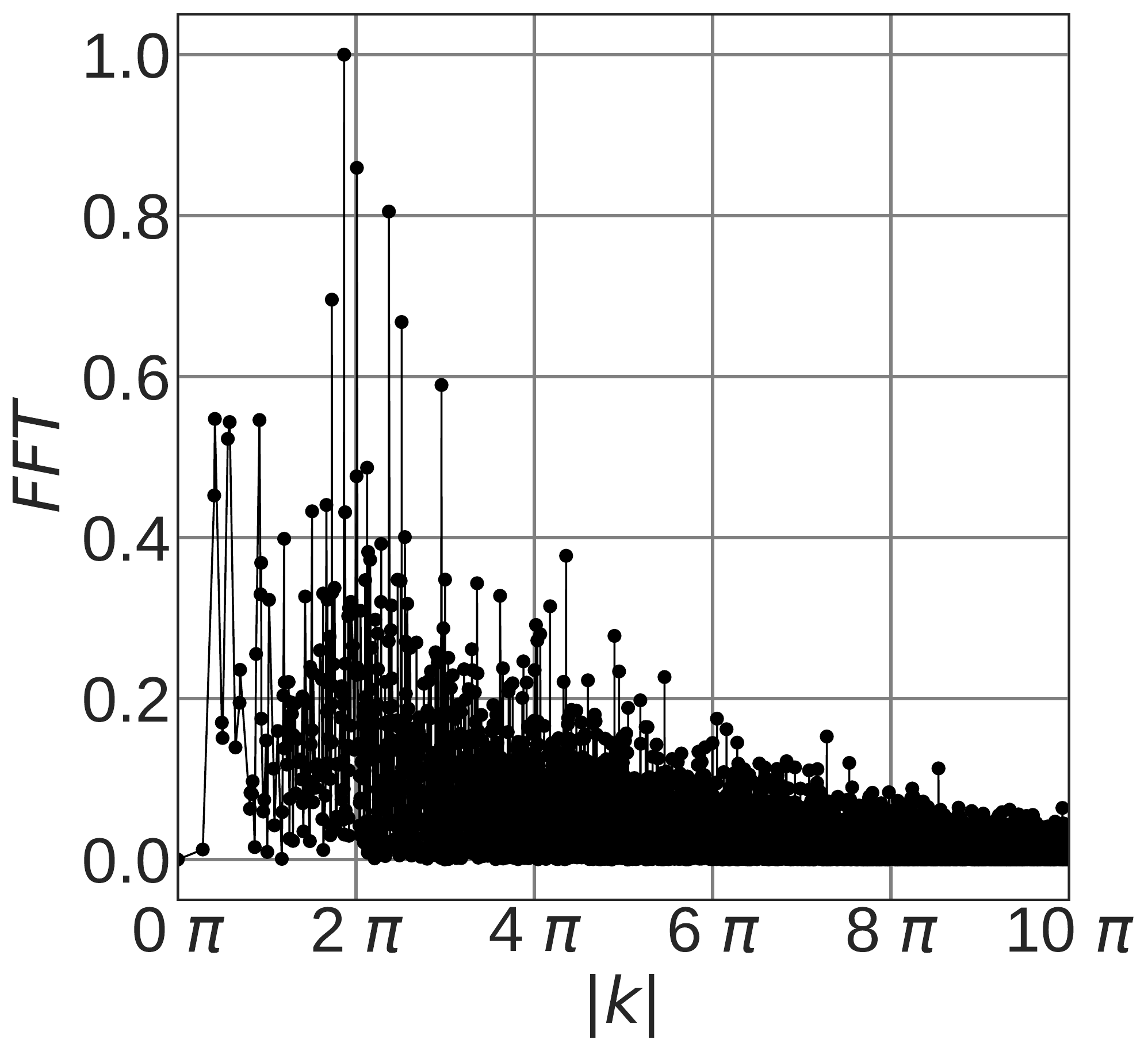}}

\caption{\label{Sfig:fft_all}
Amplitude spectrum of the fast fourier transform (FFT) of the scalar field associated to simulations (a,b) and CT-scans of real nests (c,d) shown at the top of figure 
4. 
The local maxima correspond to the emergent characteristic length scales which are reported in figure 4.
Wave numbers $k$ are normalized by the characteristic length scale.
}
\end{figure}

In its simplified form the growth equation 
(3)
can be easily linearised around $f_0=0.5$
and solved in Fourier space with solutions of the form $f=f_0+\hat{f}\exp{[i\boldsymbol{k}\cdot\boldsymbol{x} + \sigma t]}$
which gives the following dispersion relation:
\begin{equation}\label{eq:disp_relation}
\sigma(k)=\tilde{f}dk^2-k^4
\end{equation}
where for simplicity we wrote $|\boldsymbol{k}|=k$ and $\tilde{f}=f_0(1-f_0)$. 
The dispersion relation is shown in figure \sref{sfig:disp_relation}.
As $d$ is strictly positive $\sigma(k)$ changes sign in $k_0=\tilde{f}d)^{1/2}$ and has a local maximum at $k_{max}=(\tilde{f}d/2)^{1/2}$: equation 
(3)
is linearly unstable with the most unstable length scale at $\lambda_{max}=2\pi/k_{max}$ while all the small scale disturbances beyond the cutoff length $\lambda_0=2\pi/k_{0}$ are damped.
Note that $k_0>0$ independently of $d$, thus the system is formally always unstable. 
Nonetheless a threshold is fixed by the smallest $k$ which can appear in a given domain of size $L$ which is $k=2\pi/L$. 
The instability threshold $d_c$ is then fixed by $L$ the size of the domain, with $d_c=(2\pi/L)^2/\tilde{f}$ the marginal condition. 
In figure \sref{Sfig:fft_all} we report the fast Fourier transform performed on the scalar field associated to two simulations and the computer-tomography images of two nests collected from the field. One observes the presence of a local maximum which denotes the emergence of a characteristic length scale.

\section{Segmentation and curvature statistical distribution}\label{ssec:curvature_distribution}
\begin{figure}
\hfill
\centering
\subfigure{\includegraphics[height=0.25\linewidth]{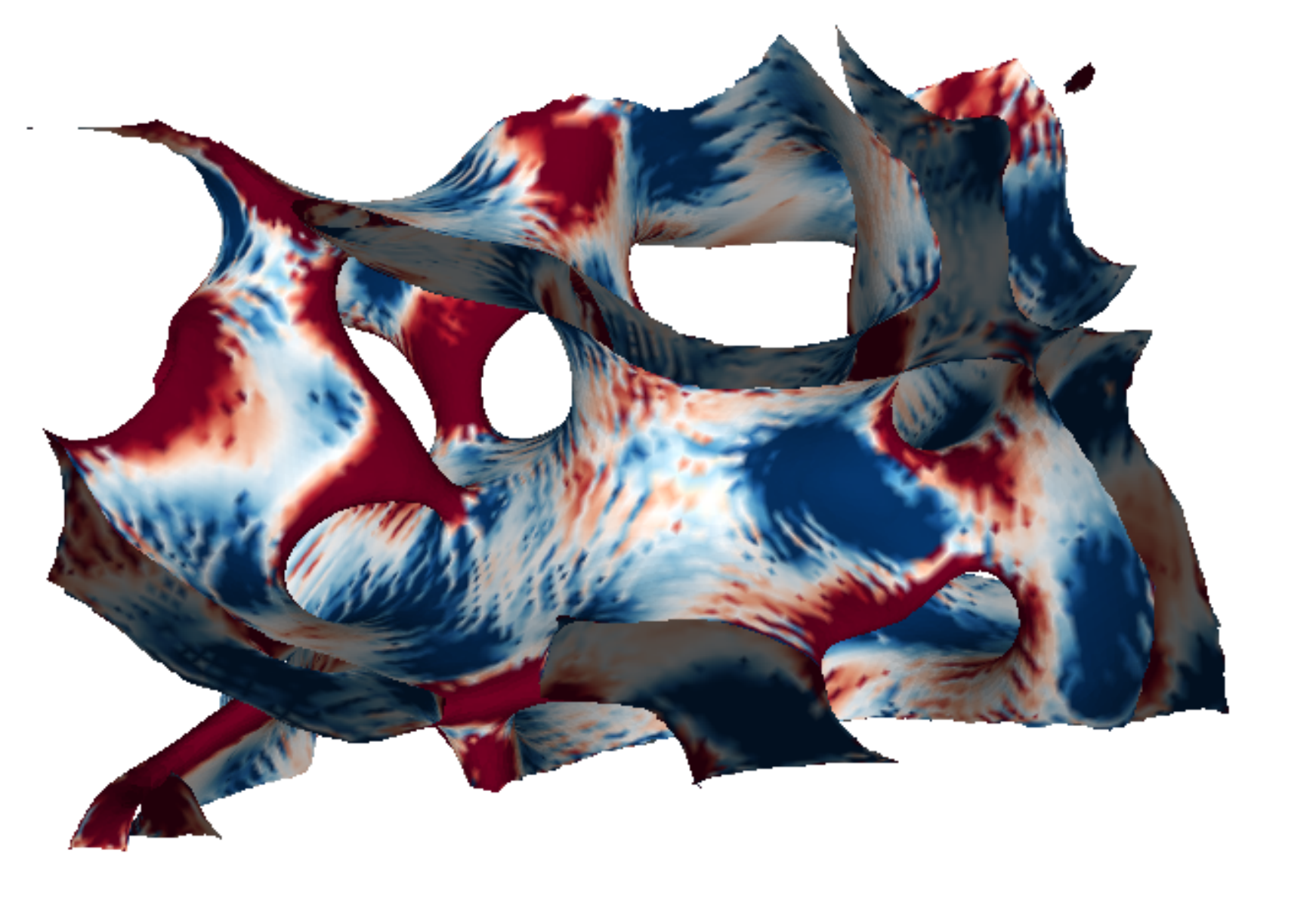}}
\hfill
\subfigure{\includegraphics[height=0.25\linewidth]{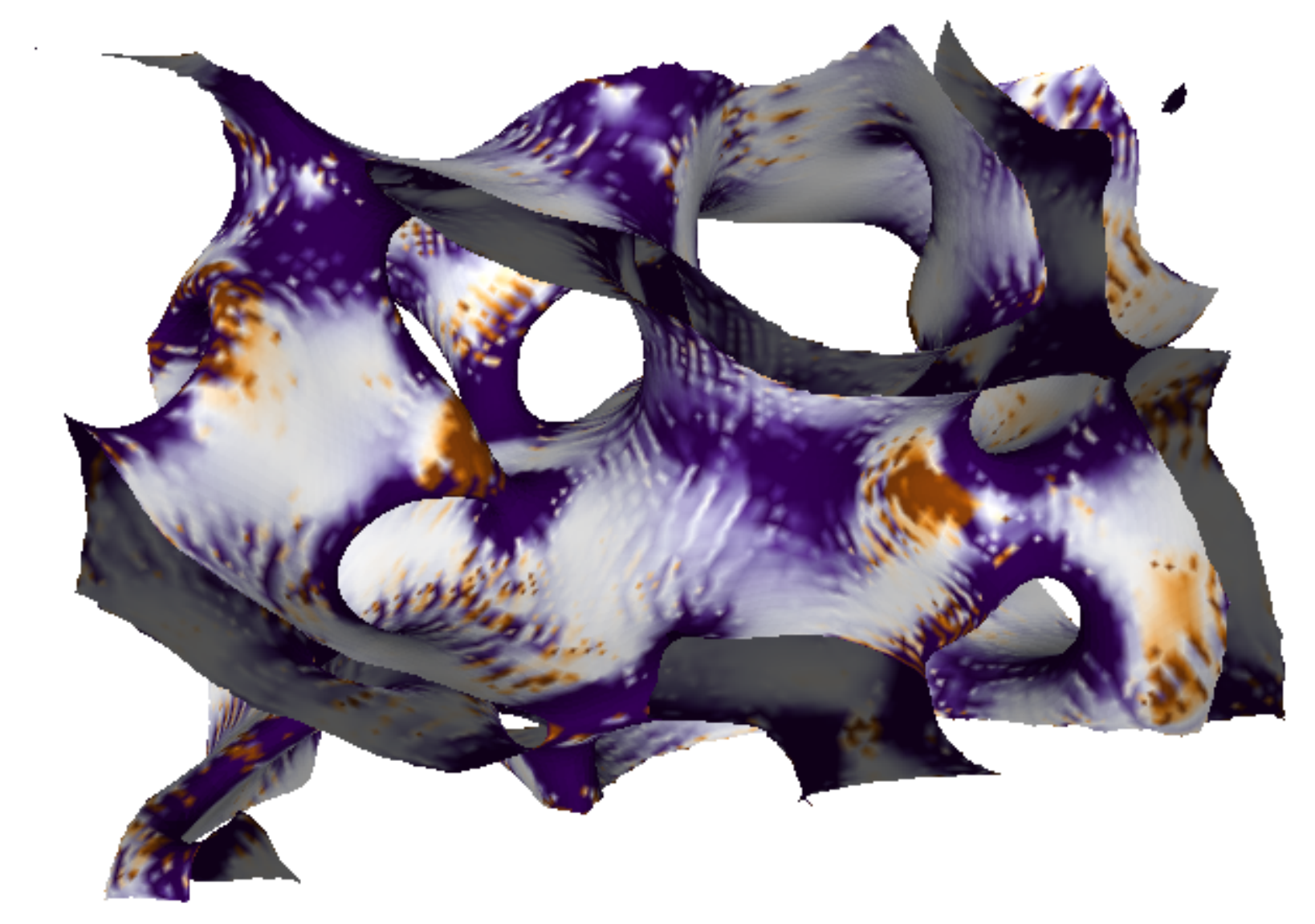}}
\hfill
\centering
\subfigure{
\begin{overpic}[height=0.105\linewidth]{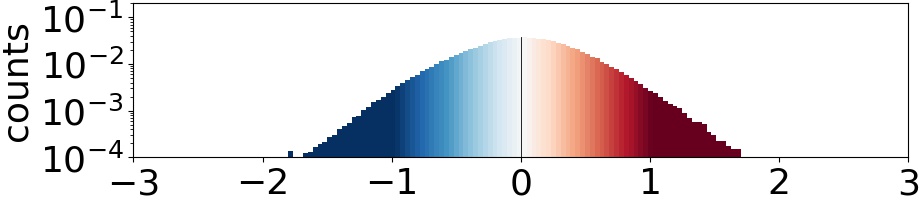}
\put(-9,15) {(a)}
\end{overpic}
}
\subfigure{\includegraphics[height=0.105\linewidth]{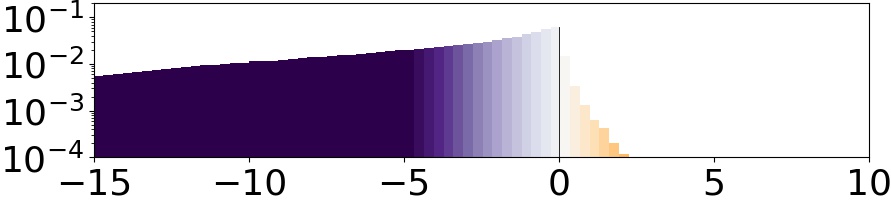}}
\subfigure{
\begin{overpic}[height=0.105\linewidth]{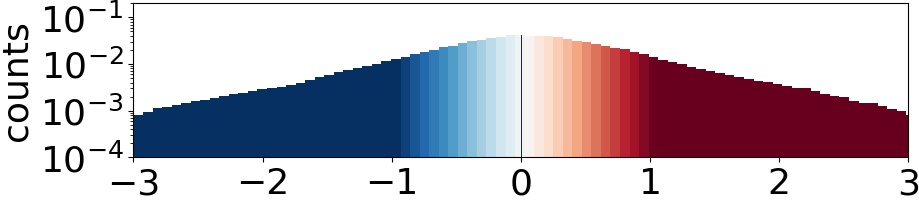}
\put(-9,15) {(b)}
\end{overpic}
}
\subfigure{\includegraphics[height=0.105\linewidth]{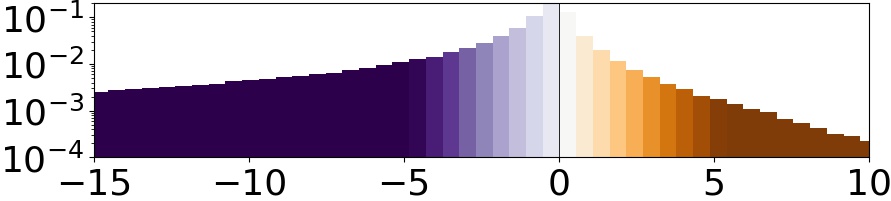}}
\subfigure{
\begin{overpic}[height=0.105\linewidth]{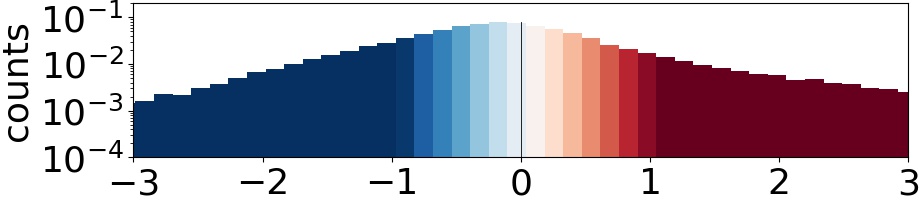}
\put(-9,15) {(c)}
\end{overpic}
}
\subfigure{\includegraphics[height=0.105\linewidth]{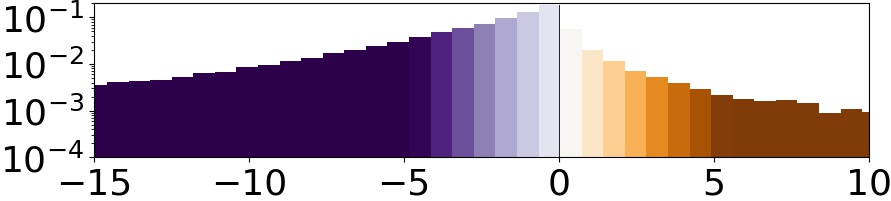}}
\subfigure{
\begin{overpic}[height=0.105\linewidth]{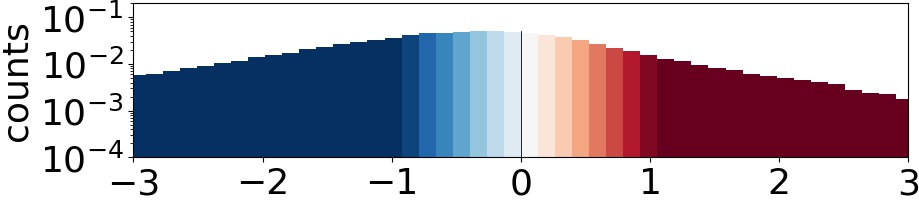}
\put(-9,15) {(d)}
\end{overpic}
}
\subfigure{\includegraphics[height=0.105\linewidth]{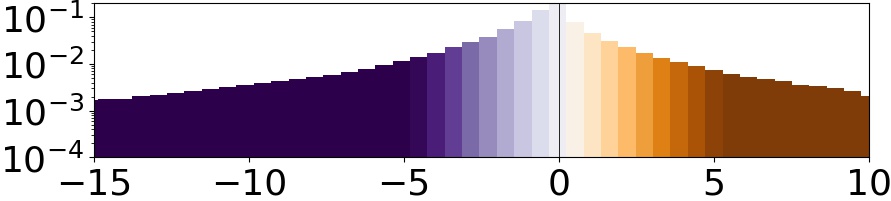}}
\subfigure{
\begin{overpic}[height=0.105\linewidth]{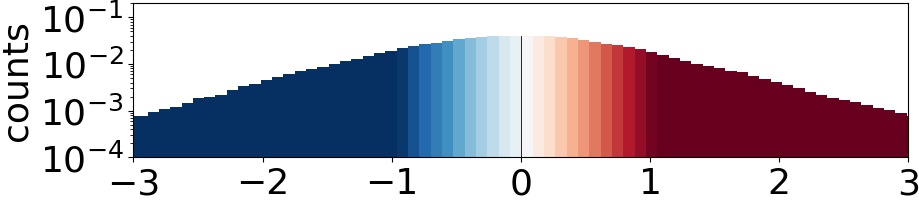}
\put(-9,15) {(e)}
\end{overpic}
}
\subfigure{\includegraphics[height=0.105\linewidth]{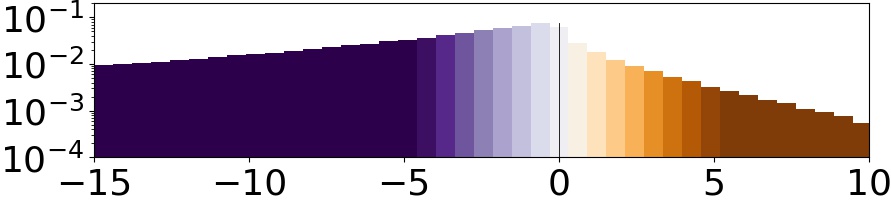}}
\subfigure{
\hfill
\begin{overpic}[height=0.105\linewidth]{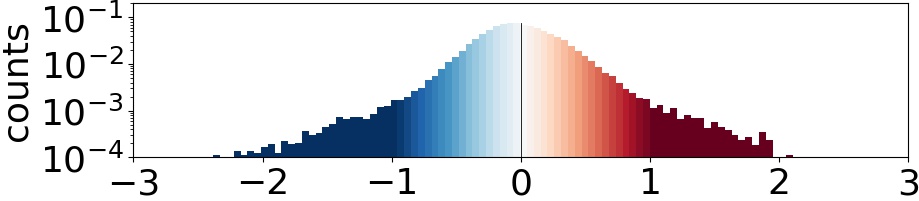}
\put(-9,15) {(f)}
\put(55,-7) {$H$}
\end{overpic}
}
\subfigure{
\begin{overpic}[height=0.105\linewidth]{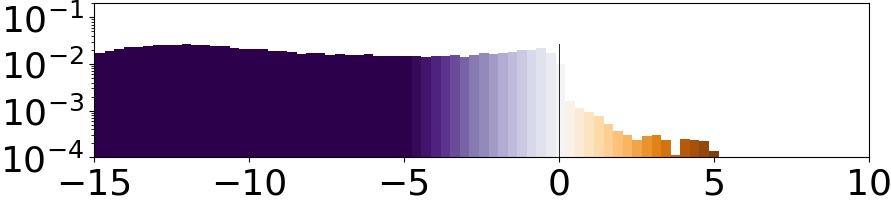}
\put(60,-7) {$\Gamma$}
\end{overpic}
}
\vspace{1em}
\caption{\label{Sfig:HG_counts}
Normalised frequency of mean curvature $H$ (left) and Gaussian curvature $\Gamma$ (right) for different surfaces, (a): simulation initiated with white noise,
(b): simulation initiated with a nest-like seed, (c): nest fragment from Guyane-\textit{N.ephratae}, (d): nest fragment from New South Wales (AUS)-\textit{N. walkeri}, (e): pass-band filtered noise, (f): gyroid minimal surface. The 3D rendering at the top represent the surface of nest (c) coloured according to the local value of $H$ (left) and $\Gamma$ (right).}
\end{figure}

In this section we explain how we obtain surfaces from a scalar field and construct the curvature distributions shown in figure \sref{Sfig:HG_counts} and 
6.
The protocol includes five steps: (i) smoothing and segmentation of the original scalar field, (ii) isosurface extraction, (iii) smoothing of the surface and (iv) computation of curvature, (v) statistical analysis of curvature distribution.

The first step only concerns nest samples collected in the field, in fact CT images must be processed to remove some of the small scale roughness of the nests and the noise introduced by the CT-scanner.
First, we filter the original greyscale images by eliminating 10 to 20\% of the highest Fourier components, which conveniently reduces the size of our data and removes isolated noise voxels due to acquisition noise or small spurious particles inside the nests (dust, dead termites, etc.). We follow this operation by applying a Gaussian filter ($\sigma=1$), mainly to remove the ripples introduced by the previous sharp low-pass filtering operation. After this initial blurring operation, we binarise our stack of images and we apply a second Gaussian filter ($\sigma=1$). 

This final smoothing operation allows having a smooth transition between walls and galleries while having a uniform value $f=1$ deep in the walls and $f=0$ in the middle of a gallery. Note that the threshold for segmentation must be chosen by hand and adjusted in order to preserve the topology of the original scanner which is inspected comparing slices of the original stack to the processed ones. 
\noindent

The second step is the isosurface extraction, that results in the production of a three-dimensional triangular mesh which represent the iso-contour surface of the scalar field at a given threshold $f_0$. We use the efficient Lewiner implementation of the Marching Cubes algorithm \citep{Lewiner2003} provided by Python scikit-image library \citep{scikit-image}. This essentially performs an iteration over all the voxels of the considered volume in which a new surface element is drawn for all the voxels where there is at least one vertex with $f>f_0$ and one with $f<f_0$.

\begin{figure}
\centering
\subfigure{\includegraphics[width=0.4\linewidth]{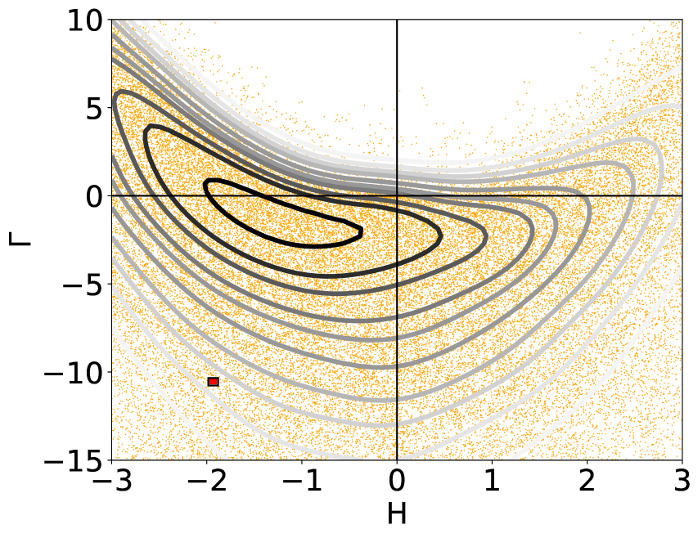}}
\subfigure{\includegraphics[width=0.4\linewidth]{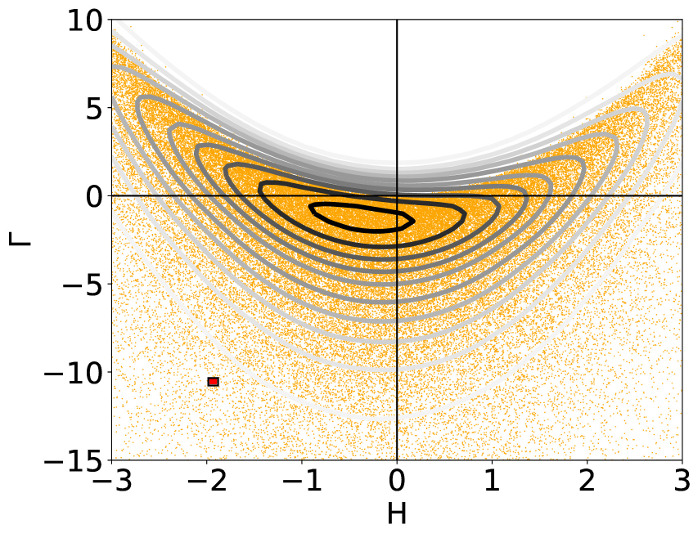}}

\subfigure{\includegraphics[width=0.4\linewidth]{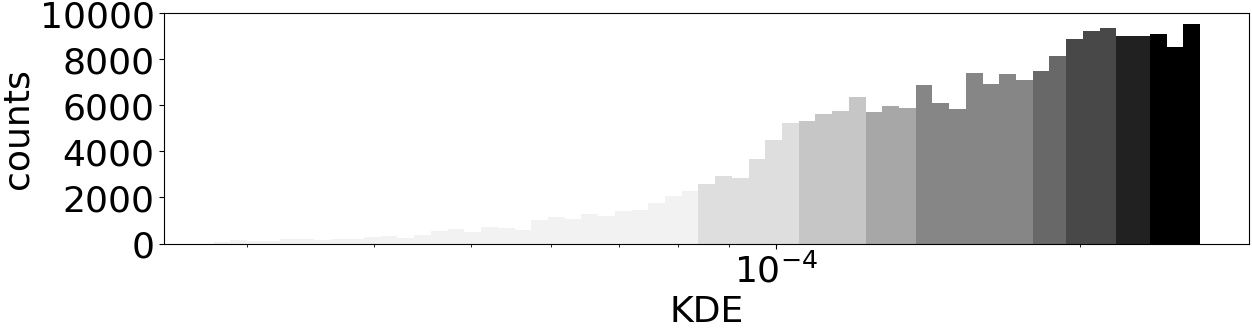}}
\subfigure{\includegraphics[width=0.4\linewidth]{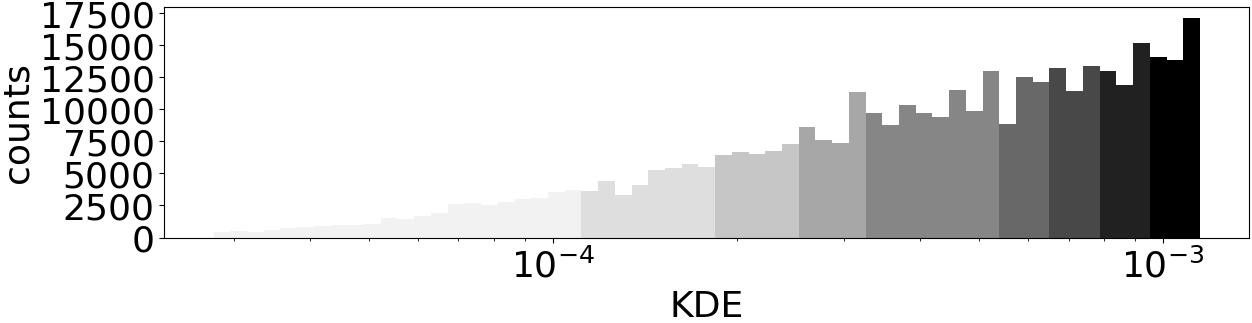}}

\subfigure{\includegraphics[width=0.4\linewidth]{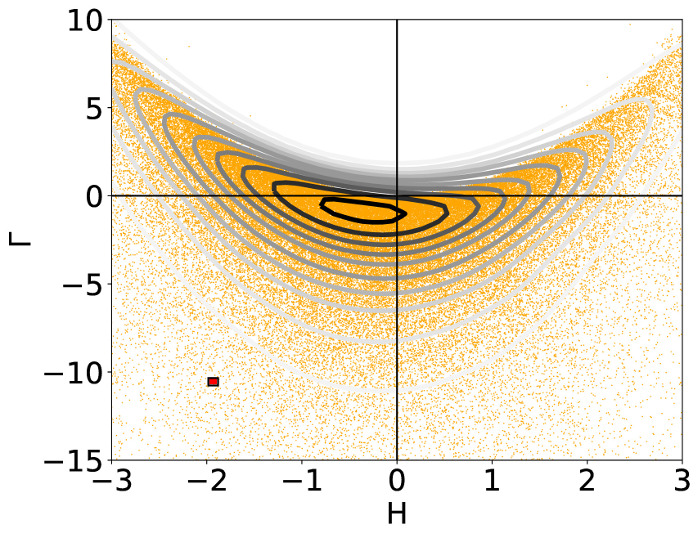}}
\subfigure{\includegraphics[width=0.4\linewidth]{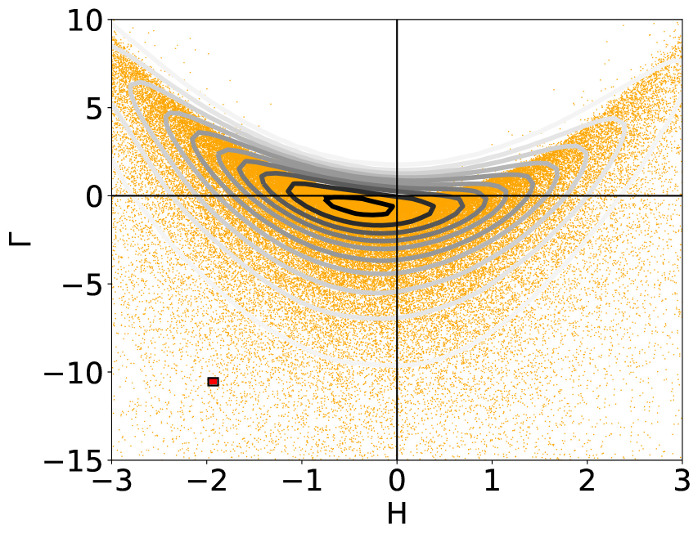}}

\subfigure{\includegraphics[width=0.4\linewidth]{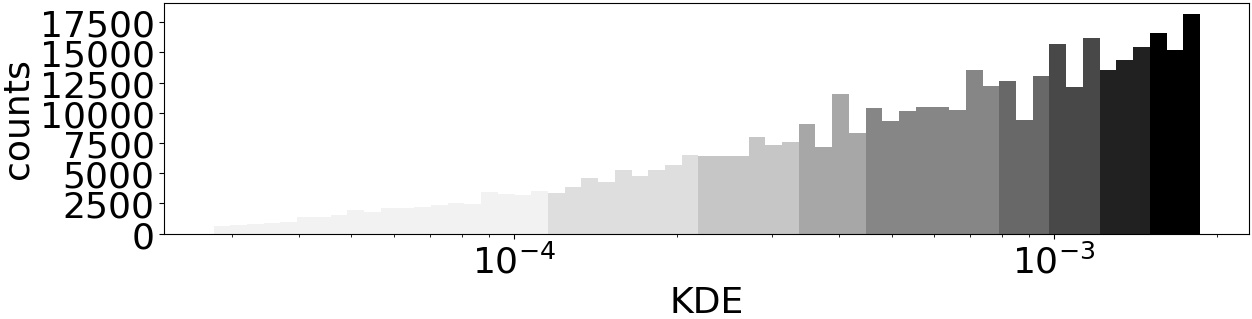}}
\subfigure{\includegraphics[width=0.4\linewidth]{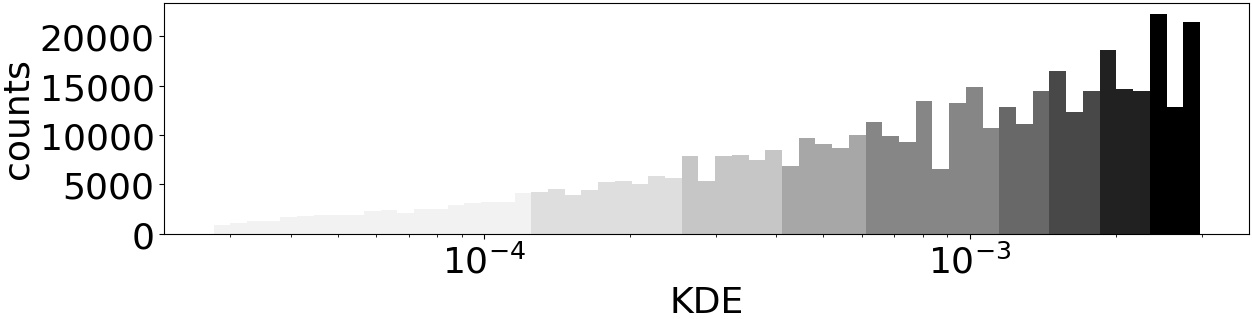}}
\caption{\label{sfig:scan_46_lap}
Curvature distribution for a \textit{Nasutitermes walkeri} nest sample, where Laplacian smoothing algorithm was applied 0, 5, 10, and 20 times. The distribution remains relatively stable after 10 iterations.}
\end{figure}

\begin{figure}
\centering
\subfigure{\includegraphics[width=0.4\linewidth]{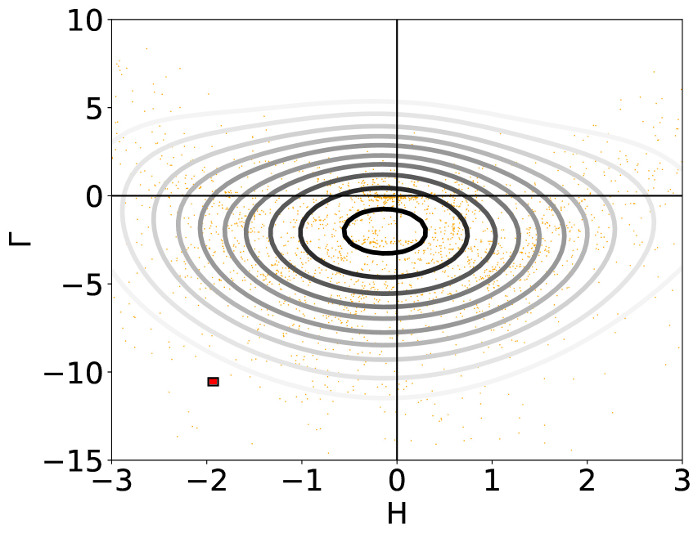}}
\subfigure{\includegraphics[width=0.4\linewidth]{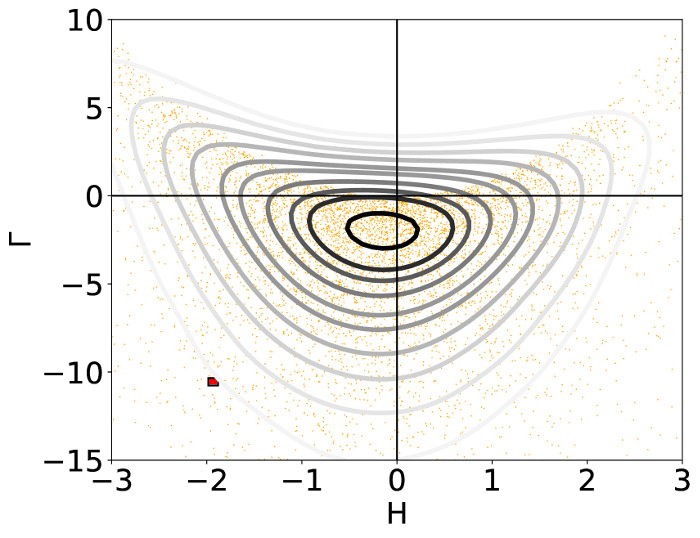}}

\subfigure{\includegraphics[width=0.4\linewidth]{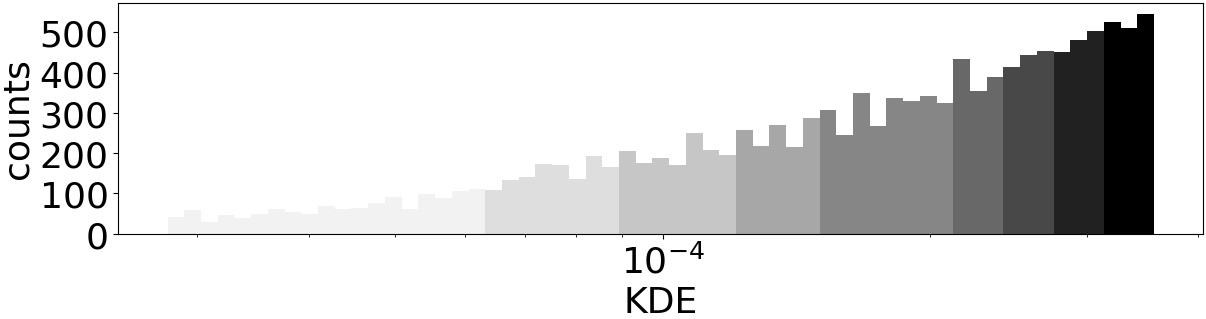}}
\subfigure{\includegraphics[width=0.4\linewidth]{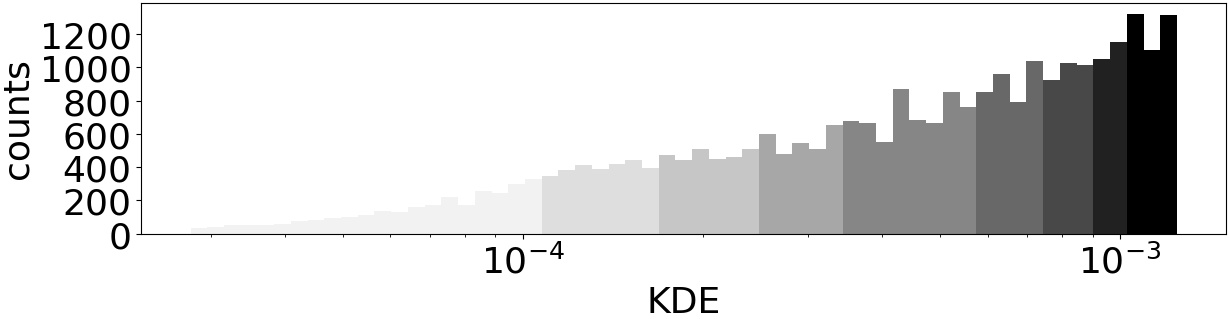}}

\subfigure{\includegraphics[width=0.4\linewidth]{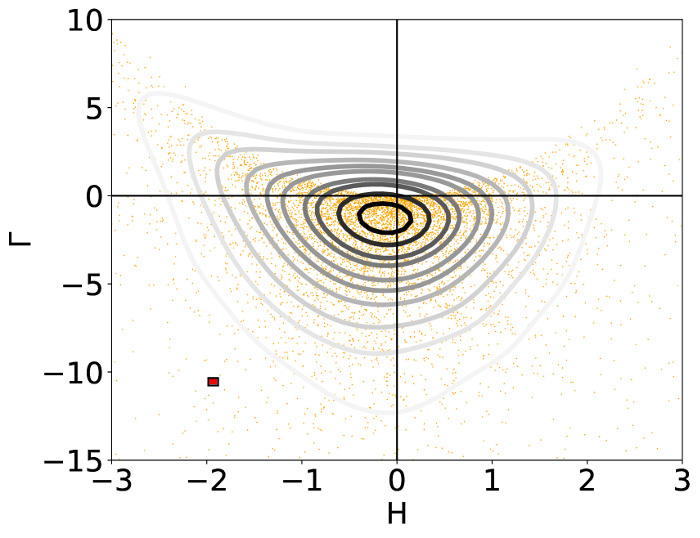}}
\subfigure{\includegraphics[width=0.4\linewidth]{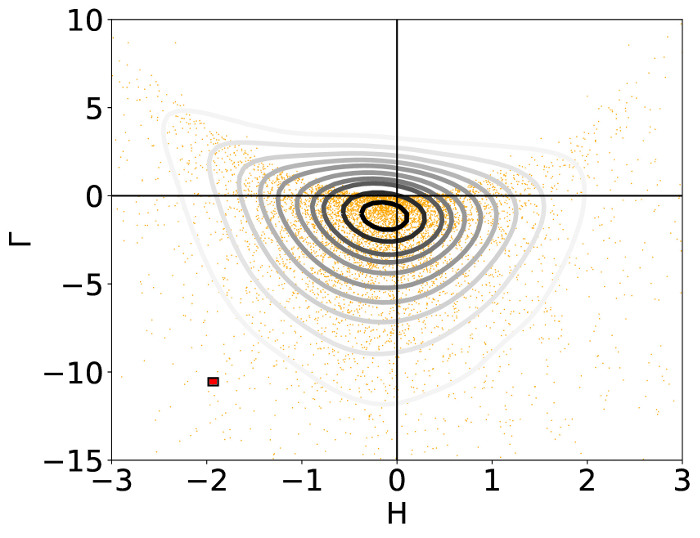}}

\subfigure{\includegraphics[width=0.4\linewidth]{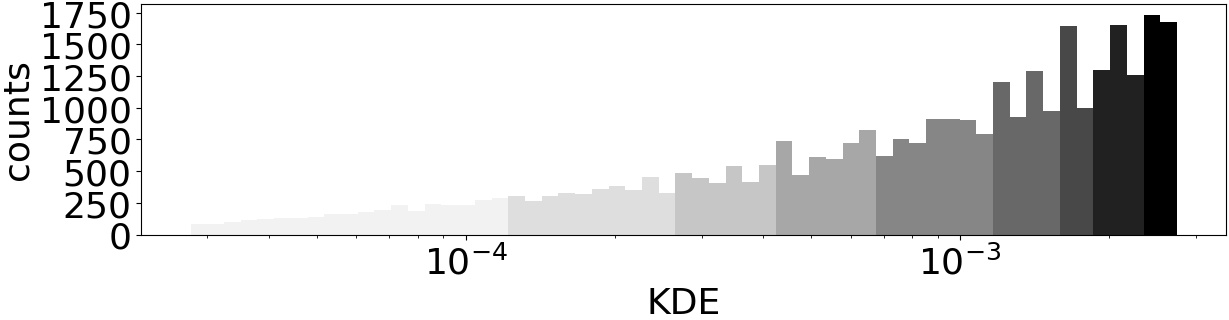}}
\subfigure{\includegraphics[width=0.4\linewidth]{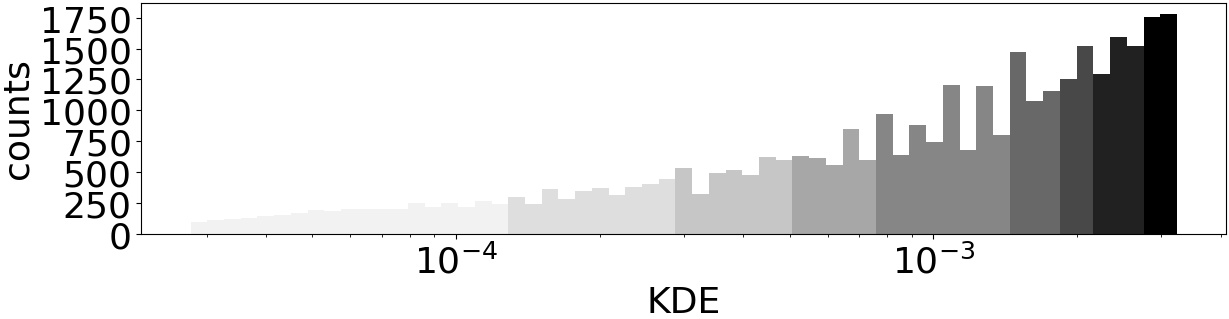}}
\caption{\label{sfig:scan_5100_lap}
Curvature distribution for a \textit{Nasutitermes ephratae} a sample, where Laplacian smoothing algorithm was applied 
0, 5, 10, and 20 times. One observes that the distribution remains relatively stable after 10 iterations.}
\end{figure}

\begin{figure}
\centering
\subfigure{\includegraphics[width=0.4\linewidth]{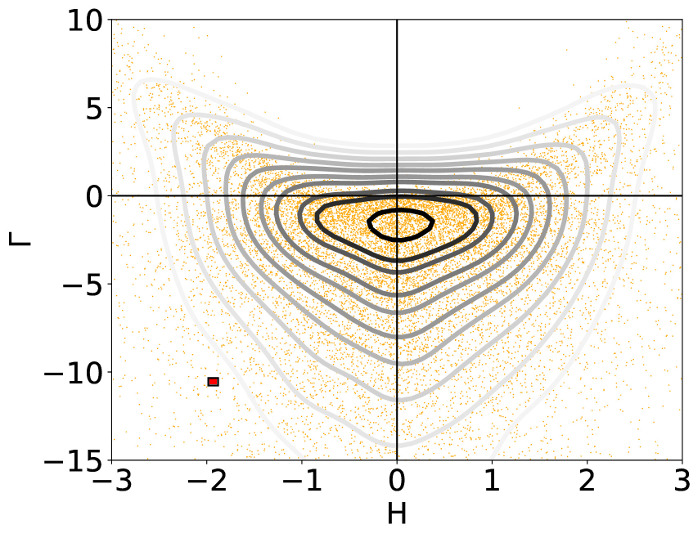}}
\subfigure{\includegraphics[width=0.4\linewidth]{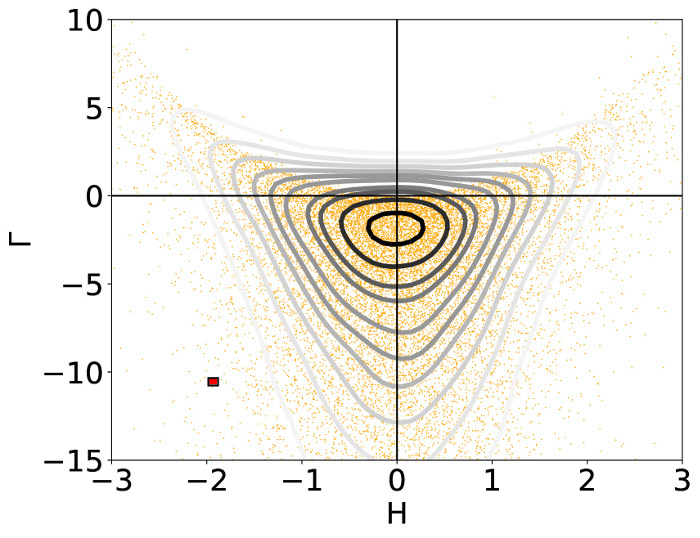}}

\subfigure{\includegraphics[width=0.4\linewidth]{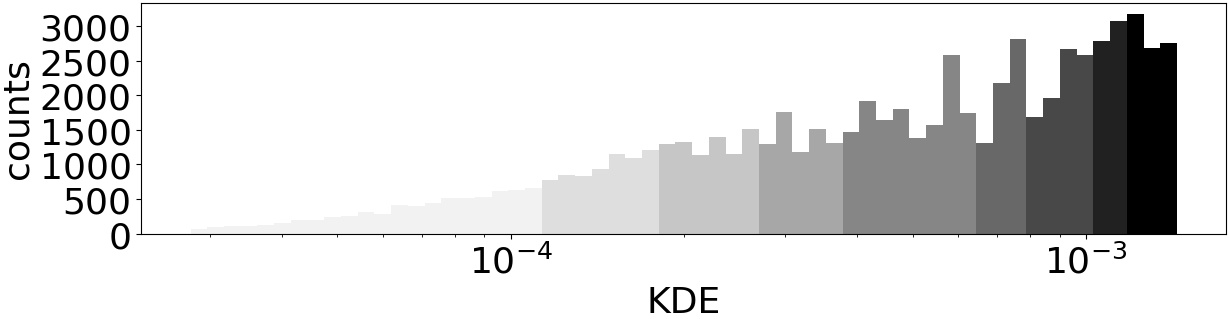}}
\subfigure{\includegraphics[width=0.4\linewidth]{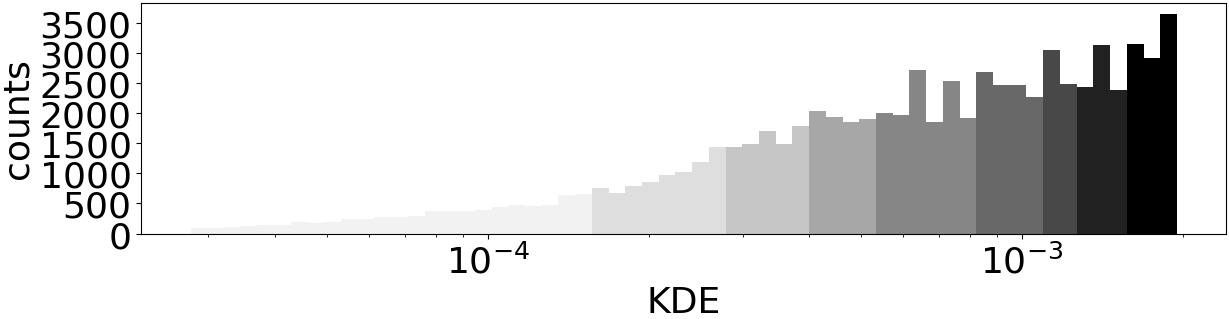}}

\subfigure{\includegraphics[width=0.4\linewidth]{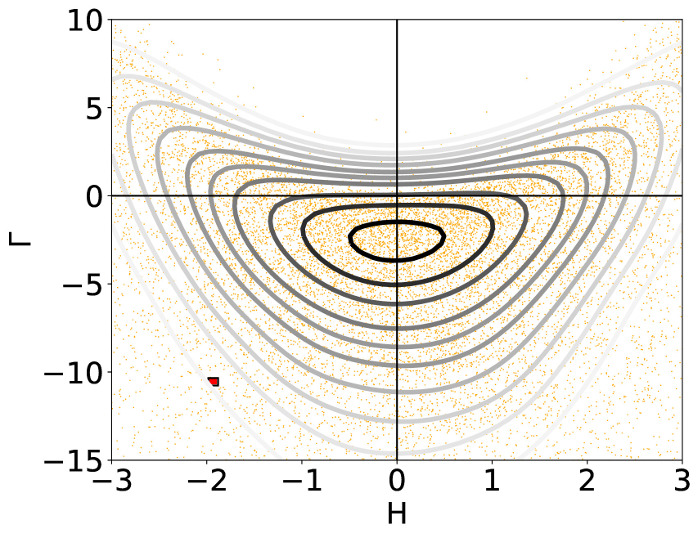}}
\subfigure{\includegraphics[width=0.4\linewidth]{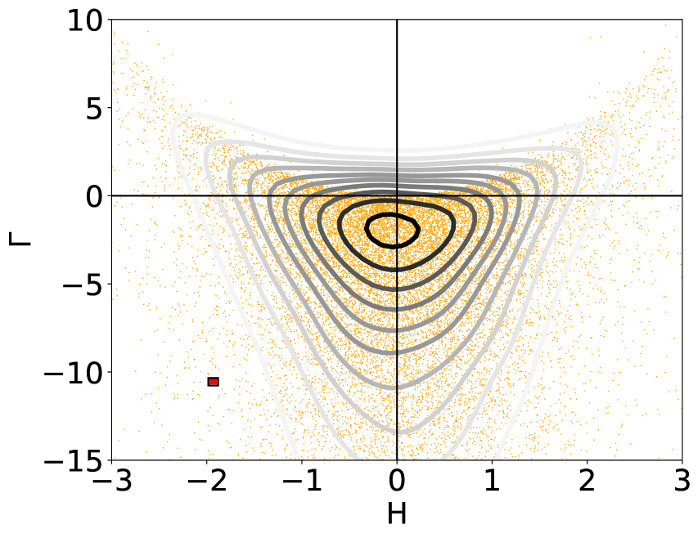}}

\subfigure{\includegraphics[width=0.4\linewidth]{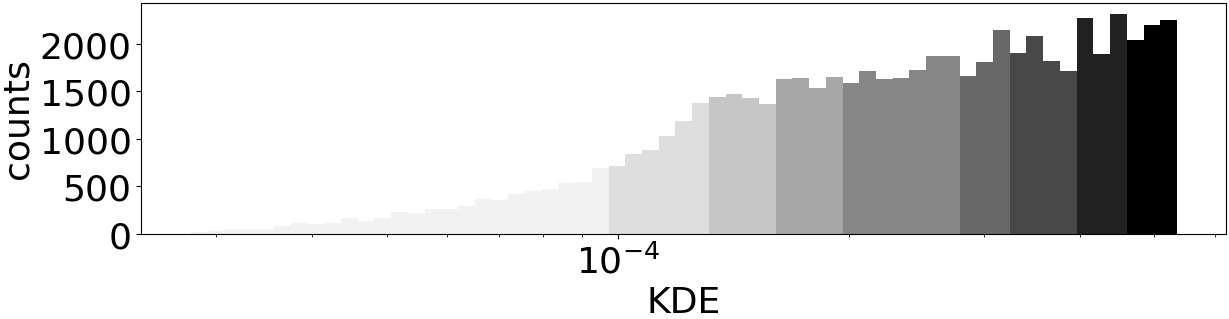}}
\subfigure{\includegraphics[width=0.4\linewidth]{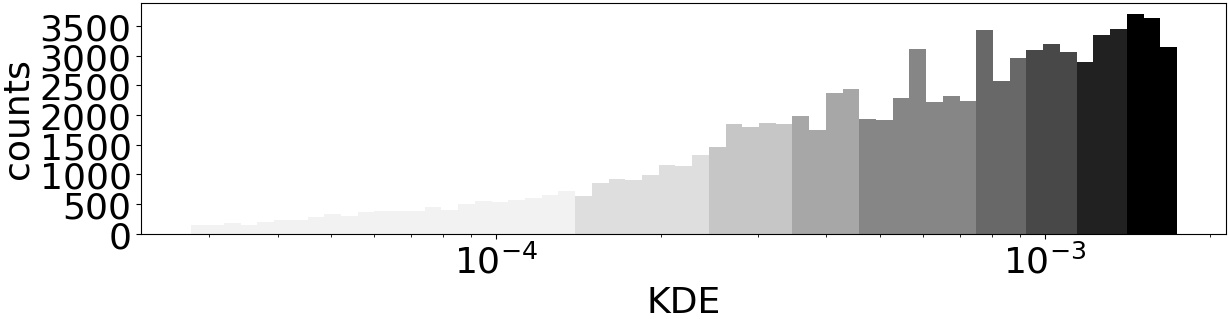}}
\caption{\label{sfig:noise_lap}
Top: Curvature distribution for synthetic noise where Laplacian smoothing algorithm was applied for 0 (left) and 20 (right) iterations.
Bottom: Same as above but a white noise was added to the initial surface. The addition of noise initially introduces a significant change in the curvature distribution (left top and bottom) but this can easily be removed by applying the Laplacian smoothing algorithm.
}
\end{figure}

As a third step we compute the mean curvature $H$ and the Gaussian curvature $\Gamma$ at each vertex of the mesh using the algorithm of \cite{Meyer2003}. Mean curvature basically corresponds to the sum of all edges connecting the considered vertex to its neighbors, while Gaussian curvature is essentially the sum of the angles between these edges and those connecting the neighbor vertices weighted by the area of
the corresponding face. Finally both $H$ and $\Gamma$ are averaged over the three vertices of a surface element. 
Prior to this computation the coordinates of all vertices are normalised by the typical length given by the FFT of the original scalar field: with this choice a surface element whose curvature is $(H,\Gamma)=(1,1)$ corresponds to an element of spherical surface of radius equal to the typical distance between walls. 
In addition, as we are using a first-neighbors algorithm, we expect the curvature computation to be extremely sensitive to the mesh roughness which results from the discrete nature of the original scalar field (sampled in discrete voxels).
 In order to remove this small scale noise (which is not relevant to the actual characteristic scale of the structure), each triangular mesh was smoothed replacing each vertex coordinates with the average of the coordinates of neighboring vertices.
This step is crucial, since on one side we want to get rid of this sensitivity and on the other side we ultimately want surface curvature to characterize our objects, so that we must avoid modifying the surface too much.
Nonetheless, we show that, by repeating the smoothing operation several times, curvature properties seem to have converged after at most $20$ iterations, as it is shown in figure \sref{sfig:scan_46_lap} and \sref{sfig:scan_5100_lap}.
Moreover, we observe that if we add spurious small scale noise to a smoothed mesh, the curvature distribution significantly changes, as it is shown in figure \sref{sfig:noise_lap}.

The last steps consist in producing the curvature distribution of the analysed surfaces. 
In figure \sref{Sfig:HG_counts} we report the normalised frequency of occurrence for $H$ and $\Gamma$ in the form of histograms for a sample of two different simulations (a,b), two different nest fragments (c,d), and two benchmark synthetic surfaces (e,f) where (e) is a random surface obtained by segmentation of a volume of band-pass filtered white noise and (f) is a minimal surface (i.e. $H=0$ everywhere) called gyroid.
Subsequently, we present the same data in the two dimensional diagram $(H,\Gamma)$ and estimate the probability density $\rho_{kde}$ with a Gaussian kernel which is defined as below:
\begin{equation}\label{seq:gaussian_kde}
\rho_{kde}(\boldsymbol{x})=\frac{1}{nh}\sum\limits_{i=1}^n e^{\frac{|\boldsymbol{x}-\boldsymbol{x}_i|}{2h^2}},
\end{equation}
where $\boldsymbol{x}$ is one generic point in the plane $(H,\Gamma)$ while $\boldsymbol{x}_{i=1,n}$ are all the surface elements.   
This tool is provided by the Python scipy library and the bandwidth $h$ is estimated with the rule-of-thumb \citep{Scott1992}.
The cumulative probability density associated with $\rho_{kde}$ is shown in figure 
6, 
\sref{sfig:scan_46_lap}, \sref{sfig:scan_5100_lap} and \sref{sfig:noise_lap} where each contour corresponds to the successive ten percentiles from the most frequent area (black) to the least frequent area (grey).
Dots correspond to the raw data before the kernel density estimation is realised. 
One observes that raw data are confined to remain below the parabola $\Gamma=H^2$ which is coherent with the definition of $H$ and $\Gamma$
given in equation 
(5). 
Note that kernel density estimation smoothens this boundary which explains why contour lines (but no dots) extend to the forbidden region.

\bibliographystyle{abbrvnat}
\bibliography{nasute}